\documentclass[fleqn,usenatbib]{mnras}

\usepackage{newtxtext,newtxmath}

\usepackage[T1]{fontenc}

\DeclareRobustCommand{\VAN}[3]{#2}
\let\VANthebibliography\thebibliography
\def\thebibliography{\DeclareRobustCommand{\VAN}[3]{##3}\VANthebibliography}

\usepackage{graphicx}	
\usepackage{amsmath}	
\usepackage{xcolor}
\usepackage{comment}
\usepackage{pdflscape}
\usepackage{tablefootnote}
\usepackage{natbib}
\usepackage{threeparttable}

\usepackage[export]{adjustbox}
\usepackage{tabularx}
\usepackage{caption}
\usepackage{subcaption}

\newcommand{\Msun}{\ensuremath{\mathrm{M}_\odot}}

\title[Determining cosmological growth parameter for stellar -
mass black holes]{Determining cosmological growth parameter for stellar - mass black holes}

\author[E. Mlinar et al.]{Ema Mlinar$^{1}$\thanks{E-mail: mlinar.ema@gmail.com}, Tomaž Zwitter$^{1}$
\\
$^{1}$Faculty of Mathematics and Physics, University of Ljubljana, Jadranska 19, 1000 Ljubljana, Slovenia
}

\date{Accepted XXX. Received YYY; in original form ZZZ}

\pubyear{2023}

\begin{document}

\defcitealias{Jenniskens1994}{JD94}
\defcitealias{HobbsYork2009}{HY09}
\defcitealias{GalazutdinovMusaev2000}{GM00}
\defcitealias{Tuairisg2000}{TC00}
\defcitealias{Weselak2000}{WS00}
\defcitealias{Fan2019}{FH19}
\defcitealias{Sonnentrucker2018}{SY18}

\label{firstpage}
\pagerange{\pageref{firstpage}--\pageref{lastpage}}
\maketitle

\begin{abstract}

It has recently been suggested that black holes (BHs) may grow with time, so that their mass is proportional to the cosmological scale factor to the power $n$, with suggested values $n \approx 3$ for supermassive BHs in elliptical galaxies. Here we test these predictions with stellar mass BHs in X-ray binaries using their masses and ages. We perform two sets of tests to assess the compatible values of $n$. First, we assume that no compact object grows over the Tolman-Oppenheimer-Volkof limit which marks the borderline between neutron stars and BHs. We show that half of BHs would be born with a mass below this limit if $n=3$ applies. The possibility that all BHs were born above the limit is rejected at $4\,\sigma$ if $n=3$ applies. In the second test, we assume that masses of BHs at their formation stay the same over cosmic history. We compare the mass distribution of the youngest BHs, which could have not grown yet, to their older counterparts. Distributions are compatible for $n = -0.9^{+1.3}_{-4.6}$, with $n=3$ excluded formally with 87\%\ confidence. This result may be biased, because massive BHs tend to have a massive companion. Correcting for this bias yields $n\approx 0$. We can therefore conclude that while our results are not a clear rejection of BH scaling with $n=3$, we show that $n=0$ is much more consistent with the data.

\end{abstract}

\begin{keywords}
stars: black holes -- (stars:) binaries (including multiple): close -- cosmology: theory -- X-rays: binaries
\end{keywords}



\section{Introduction}
Recently, there has been much interest in non-singular black holes (BH), particularly those with vacuum energy interiors. It was proposed that these black holes can couple to the expansion of the Universe \citep{croker19}. \citet{farrah23b} introduced a simple expression for describing the growth of non-singular black holes:
\begin{equation} \label{potencni}
    m=m_0 \left(\frac{a}{a_0}\right)^n
\end{equation}
which connects mass ($m_0$) and scale factor ($a_0$) at black hole formation to its mass ($m$) at a later scaling factor ($a$). Only values $-3 \leq n \leq 3$ are considered to be plausible. Initially, $n=3$ was used in the context of GEODEs - Generic Objects of Dark Energy \citep{croker19}. However, we note that \citet{cadoni23} claim that $n=1$ for any spherically symmetric object. 

The proposal has been tested for supermassive BHs in elliptical galaxies with $0 < z \lesssim 2.5$. \citet{farrah23b} derived that they obey the relation with $n\approx 3$. On the other hand, observations at higher redshifts ($4.5 \lesssim z \lesssim 7$) excluded $n\approx 3$ at confidence level $3\sigma$ \citep{lei23}.

If applicable, the relation \ref{potencni} should also hold for BHs with smaller, i.e.\ stellar, masses. Gravitational wave detection datasets from LIGO, Virgo and KAGRA detectors exclude $n\approx 3$ at a greater than $2.6\sigma$ level \citep{amendola23}. Complementary studies excluded the possibility of $n=3$ with the assumption that BHs should be born with masses larger than the Tolman–Oppenheimer–Volkoff (TOV) limit of $m_0\approx 2.2 \,\Msun$ \citep[e.g.][]{rodriguez23,andrae23}. 

Here we extend this approach to a larger sample of BHs in X-ray binaries. Apart from testing if $n=3$ is compatible with $m_0 > 2.2 \,\Msun$, we also check which values of $n$ do not require crossing of the limit. In addition, we use individual estimates of BH age to check which values of $n$ are compatible with the assumption that the mass distribution of stellar mass BHs is constant in the last about 10 Gyr.

The structure of the paper is as follows. In the next section we discuss observational inputs, i.e.\ masses and ages of BHs, with notes on individual objects elaborated in the Appendix. Section 3 describes our methodology, and section 4 presents the results of the two tests mentioned above. Section 5 focuses on possible caveats of our approach and compares our results to others in the literature. The main conclusions are outlined Section 6.

\section{Data}\label{podatki}

We focus on BHs in X-ray binaries. Most of them have their mass determined dynamically by the motion of matter in the accretion disk or motion of the companion star, as discussed in notes on individual objects (see Appendix A). We limit our sample to systems which are confirmed or at least highly likely to contain a black hole. We discuss 22 such systems. Moreover, the assumed coeval origin of both stars and, in some cases also a wider environment, allow us to infer an approximate age of each of the BHs. 

It is convenient to divide BHs into two age groups, with young BHs separated from the rest of the sample. Here we consider a BH to be young if we can judge it was formed less than 0.5~Gyr ago. Such BHs are unique in the sense that their masses could not have changed significantly, irrespective of the assumed $n$. 

\subsection{Young BHs}

There are four cases which suggest that the BH formed only recently,  
\begin{enumerate}
    \item 
    the object corresponds to a known supernova,
    \item supernova itself was not recorded, but its remnant is still visible,
    \item X-ray binary is a member of a young open cluster, and
    \item 
    the companion star is massive and therefore young.
\end{enumerate}
The last item assumes that the star, which later became a BH, is coeval with its binary companion. The BH itself can be only younger than its companion. 

Table~\ref{mlade} lists all young BHs. For each object, we list its assumed mass, why it belongs to the young BH group, and its orbital period, in all cases quoting the relevant source from the literature. Some masses of BHs were determined dynamically from the motion of the companion and, in many cases profile, fitting of spectral lines which are assumed to originate in the quiescent accretion disk, was used. Another possibility is the detection of quasi-periodic oscillations, which are assumed to have frequencies matching the orbital motion of particles close to the innermost stable circular orbit, as predicted by the general theory of relativity \citep[see][]{motta14}. 

Our analysis considers all young objects from Table~\ref{mlade} but the last one, to which we return in the Discussion section. The reason for a distinction is that this system is the only one not confirmed to contain a BH, because its low mass and uncertainty are consistent with a massive neutron star. Discussion on individual objects and adopted values of their parameters are in the Appendix \ref{A1}.

\begin{table*}
    \caption{Table of all systems/BHs that we consider young. 
    Label Simbad means that the spectral type of the companion star from this database is O or B, which implies that also the progenitor of the BH formed only recently. The last two columns list the orbital period and its source. As its error is generally much smaller than 0.1 days, we refrain from quoting the errors. Discussion on individual objects is in the Appendix \ref{A1}.}
    \begin{tabular}{lcccccc}
    \hline
    Name & BH mass/$M_\odot$ & Reference & Why young? & Reference & Period/d & Reference\\
    \hline
    GRO J1655--40 & $5.31\pm 0.07$ & 1 & open cluster/ & 2 & 2.6 & 3\\
     &  &  & metallicity &  &   & \\
    NGC 300 X--1 & $17\pm 4$ & 4 & companion & 5 & 1.4 & 6\\
    Cyg X--1 & $19.2\pm 1.9$ & 7 & companion & Simbad     & 5.6 & 8\\
    VFTS 243 & $10.1\pm 2$ & 9& companion & Simbad & 10.4 & 10\\
    M33 X--7 & $15\pm 1.45$ & 11& companion & 12 & 3.5 & 13\\
    LMC X--1 & $10.14 \pm 0.65$ & 14 & companion & Simbad & 3.9 & 15\\
    V4641 Sgr & $6.4\pm 0.6$ & 16 & companion & Simbad & 2.8 & 17\\
    LMC X--3 & $6.98 \pm 0.56$ &18 & companion & Simbad & 1.7 & 19\\
    MAXI J1535--571 & $10 \pm 2$ & 20: & supernova rmn& 21 & $\sim 5$ & 22 \\
    SS 433 & $15\pm 2$ & 23 & supernova rmn & 24 & 13.1 &23\\
    \hline
    SN 1997D & about 3 & 25 & supernova & 25 & / & / \\
    \hline
    \end{tabular}
    \begin{tabular}{l}
    {\bf References:}
    1: \citet{motta14};
    2:  \citet{combi07};
    3: \citet{greene01};
    4: \citet{binder21};
    5: \citet{crowther10};
    6: \citet{carpano07};\\
    7: \citet{orosz11a};
    8: \citet{brocksopp99};
    9: \citet{shenar22};
    10: \citet{shenar22};
    11: \citet{orosz07};
    12: \citet{laycock15};\\
    13: \citet{pietsch06};
    14: \citet{bhuvana21};
    15: \citet{orosz09};
    16: \citet{macdonald14};
    17: \citet{orosz01};
    18: \citet{orosz14};\\
    19: \citet{boyd01};
    20: \citet{sridhar19};
    21: \citet{sridhar19};
    22: \citet{bhargava19};
    23: \citet{bowler18};
    24: \citet{kayama22};\\
    25: \citet{benetti01}.
    \end{tabular}
    \label{mlade}
\end{table*}

\subsection{Other BHs}\label{ostale}
 
BH binaries believed to be older than 0.5~Gyr are listed in Table~\ref{stare}. The list includes 11 objects: 9 of them are X-ray binaries, and 2 are detached systems. Any objects for which only the lower limit of the mass of the compact object was determined were excluded. Examples are two objects analyzed by \citet{rodriguez23}. For each system, we list the BH's mass and the system's age and metallicity, in all cases with their errors and source. Orbital periods indicate that all objects are close binaries, except for Gaia BH1 and Gaia BH2, which are wide detached binaries.  

Discussion on masses of BHs in individual objects is in Appendix \ref{A2}. Here we discuss the age of the BH in each binary. The ages of detached systems Gaia BH1 and Gaia BH2 were determined by fitting the broadband spectral energy distribution of the companion and inferring its age from isochrone fitting \citep{andrae23}. The age of  XTE J1118--480 follows from stellar evolution calculations of its companion with its turn-off age and the time which was needed for the orbit to tighten to its present state \citep{gualandris05}. The orbital period of MAXI J1659--152 implies that it is a very close binary with orbital separation of $\gtrsim 1.33$~R$_\odot$ \citep{kuulkers13}. So the companion of the BH, which is now an about $0.2$~R$_\odot$ M5 dwarf star with a mass of about $0.20$~M$_\odot$, 
started its life with a much higher mass of 1.5~M$_\odot$. The mass transfer took time, so \citet{kuulkers13} estimate the system's present age to be 5--6 Gyr. The state of the companion star is also a clue to the age of GS 1354--64. \citet{casares04} detect its spectral lines and show it is a G0 to G5 giant. Their analysis favours a G subgiant that has evolved off the main sequence to fill its Roche lobe. This implies a relatively young age of about $2 \pm 1$~Gyr. The companion of the BH of A0620--00 had a massive progenitor of about $14$~M$_\odot$ \citep{casares17}, implying the young age of the system, even when allowing for orbital shrinking to the present 7.75hr orbital period. \citet{bartolomeo23} present a detailed computation of evolution of V404~Cyg to its current state. We use their age estimate, but with larger errors, which reflect the fact that their evolutionary scenario may not be unique and that their model does not reproduce the suggested high spin rate of the BH. Note that the supersolar metallicity of the companion star $[\mathrm{Fe}/\mathrm{H}] = 0.23 \pm 0.19$ \citep{gonzalez11} is consistent with a moderate age of this system.

Another way to estimate the age is from the metallicity of the companion. By 'metallicity' we refer to logarithmic abundance of iron [Fe/H]. We note abundances of certain elements in the atmosphere of the companion may have increased during the formation of a BH in a supernova explosion. So such systems appear younger than they are and their ages, derived from metallicity, may be somewhat underestimated. On the other hand, a kick from a supernova may change the properties of the Galactic orbit of the system, so it may end up in a more metal-poor environment of the thick disk or halo. We consider ages estimated from metallicity relatively uncertain, so we adopt conservative errors for their age estimates. 

\begin{table*}
    \caption{Table of all X-ray systems that we consider in the second data group. In the first column there is a name of the system/BH, and then in the second and third we list the mass of BH and its source, respectively. In the fifth column there is the age of the system. If the age from literature is unknown, there is a metallicity instead in the fourth column, and then the age is derived from it in the fifth column. Such ages are marked with brackets. Next we quote the source for the age/metallicity. The last two columns give the orbital period and its source. Its error is not given since it is negligible for our purpose.}
    \begin{tabular}{ cccccccc }
    \hline
    Name & BH mass/$M_\odot$ & Reference & Metallicity & Age/Gyr & Reference & Period/days & Reference\\
    \hline
    Gaia BH1 & $9.32^{+0.22}_{-0.21}$ & 1 &$-0.20 \pm 0.05$ & $7.1^{+2.0}_{-1.7}$ & 2 & 186 & 3\\
    Gaia BH2 & $8.9\pm 0.3$ & 4 &$-0.22\pm 0.02$  & $7.9^{+3.2}_{-2.7}$ & 29 & 1277 & 4\\
    XTE J1118--480 & $8.3^{+0.28}_{-0.14}$ & 5 & & $3.5 \pm 1.5$ & 6 & 0.17 & 7\\
    GRS 1009--45 & $8.5\pm 1$ & 8 & $-0.17\pm 0.2$ & ($9\pm 3$) & Starhorse & 0.28 & 9\\
    MAXI J1659--152 & $6^{+1.8}_{-1.3}$ & 10 & & $5.5 \pm 0.5$ & 11 & 0.10 & 12\\
    MAXI J1820+070 & $10.3 \pm 3.6$ & 13 & $-0.55 \pm 0.17$ & ($10\pm 3$) & 14 & 0.69 &15\\
    4U 1543--47(5) & $9.4\pm	1.0$ & 16 & $-0.11\pm 0.1$ & ($7\pm 4$) & Starhorse & 1.1 & 17\\
    GX 339--4 & $10.09\pm 1.81$ & 18 & $-0.12\pm 0.1$ & ($7.5\pm 4$) & Starhorse & 1.7 & 19\\
    GS 1354--64 & $10 \pm 2.5$ & 3 & & $2\pm 1$ & 20 & 2.5 & 21\\
    A0620--00 & $6.6\pm 0.25$ &22 & $+0.14 \pm 0.2 $ & ($2\pm 2$) & 23 & 0.3 & 24\\
    V404 Cyg & $9^{+0.2}_{-0.6}$ & 25 &  & $3.6\pm 2$ & 26, 27 & 6.5 & 28 \smallskip\\
    \hline
    \end{tabular}
    \begin{tabular}{l}
    {\bf References:}
    1: \citet{chakrabarti23};
    2: \citet{andrae23};
    3: \citet{el-badry23a};
    4: \citet{el-badry23b};\\
    5: \citet{gonzalez12};
    6: \citet{gualandris05};
    7: \citet{cook00};
    8: \citet{macias11};
    9: \citet{shahbaz97};\\
    10: \citet{molla16b};
    11: \citet{kuulkers13};
    12: \citet{kuulkers13};
    13: \citet{chakraborty20};
    14: \citet{mikolajewska22};\\
    15: \citet{torres19};
    16: \citet{ozel10};
    17: \citet{orosz98};
    18: \citet{sreehari19b};
    19: \citet{hynes03};
    20: \citet{casares04};\\ 
    21: \citet{casares09};
    22: \citet{cantrell10};
    23: \citet{casares17};
    24: \citet{esin00};
    25: \citet{khargharia10};
    26: \citet{casares17};\\
    27: \citet{bartolomeo23};
    28: \citet{casares19}.
    \end{tabular}
    \label{stare}
\end{table*}

For three objects in Table \ref{stare} we could not find any detailed chemical abundance determination. So we adopted the metallicity of the companion from the database StarHorse 2 - GaiaEDR3 \citep{queiroz23}. This database uses Gaia EDR3 spectrophotometric data together with Starhorse code \citep[see][]{queiroz18} to derive parameters such as distances, extinctions, effective temperatures and metallicities for field stars \citep{anders22}. StarHorse uses spectrophotometric parameters to calculate the posterior probability distribution over a given grid of stellar evolutionary models, using Galactic stellar-population priors. The errors in Table~\ref{stare} relate to the 16th and 84th percentile. The metallicity from other sources was preferred over this database because this database does not take into account characteristics specific for individual stars. The age of the companion star was determined from its metallicity with the relation of \citet[][their figure 6]{thompson18}, which is based on observations by the Gaia-ESO survey.

\begin{figure}
\centering
\begin{subfigure}[b]{\columnwidth}
  \includegraphics[width=\linewidth]{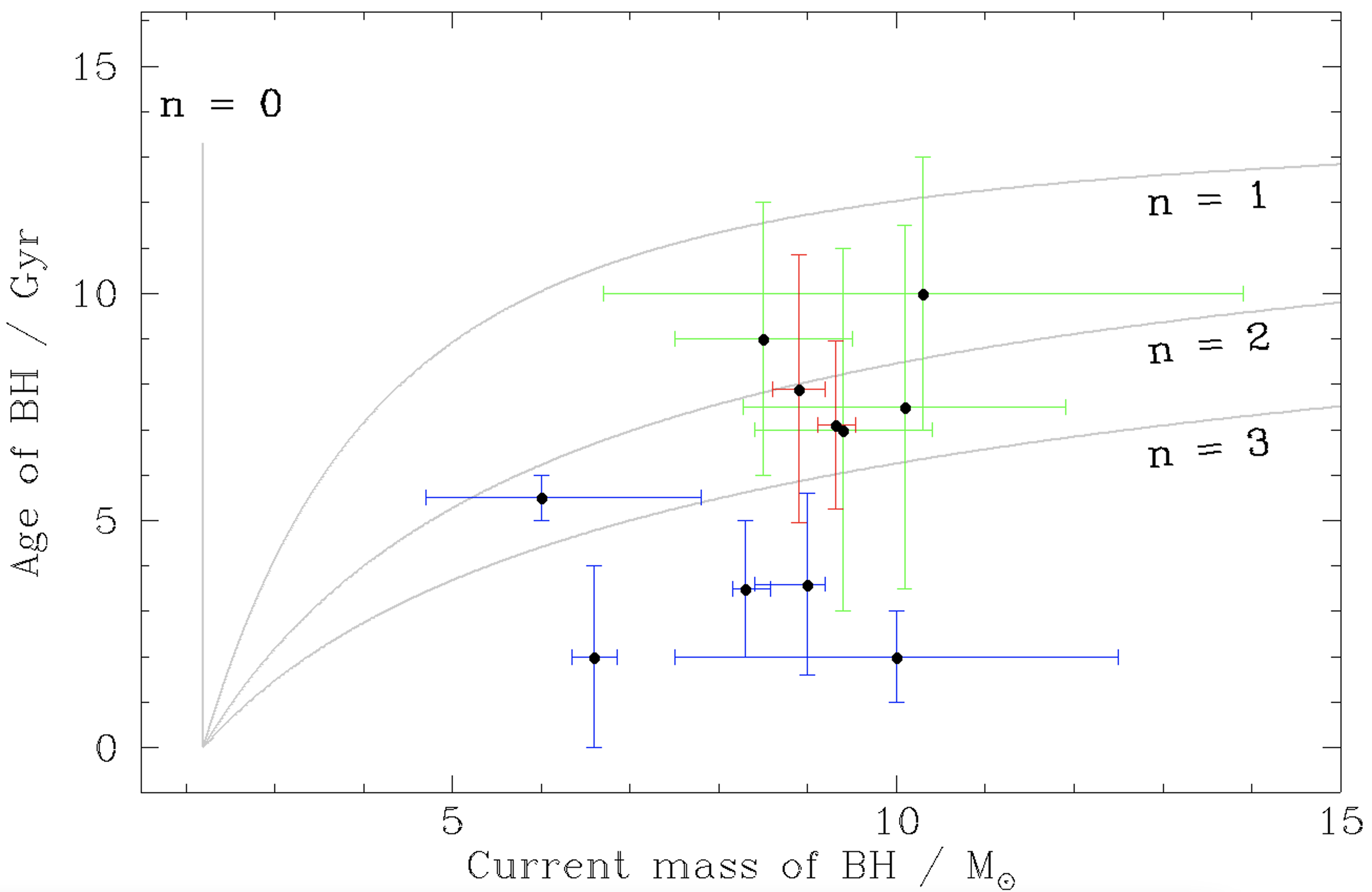}
\end{subfigure}
\caption{Relation between age and mass of BHs for objects older than 0.5~Gyr. Numerical values are quoted in Table \ref{stare}. Red errorbars mark BHs residing in wide detached binaries, Gaia~BH1 and Gaia~BH2. Blue errorbars are for X-ray binaries where the age is determined from modelling of the system's evolution, and green errorbars are for those with the age estimated from metallicity of the companion. Grey lines show the change of mass of the compact object with time, if the object is born with a mass of 2.2~M$_\odot$ and the indicated value of $n$ applies.}
\label{slikica_podatki}
\end{figure}

The age of the BH in MAXI~J1820+070 was estimated from the Galactic position of the object and an inferred mass accretion rate and evolutionary history of the system \citep{mikolajewska22}. Uncertainty of such age inference is similar to the ones judged from the StarHorse database. To separate this group of 4 objects from the rest we report their ages in Table \ref{stare} in brackets, while their masses in Figs.\ \ref{slikica_podatki} and \ref{porazdelitev mas} are plotted in green. 

Fig. \ref{slikica_podatki} reports masses and ages of BHs older than 0.5~Gyr. As already mentioned, the actual age of a BH may be slightly smaller, because a star needs some time after the formation of the binary to turn into a BH. But, considering that most BHs have rather large masses, they should have formed from massive progenitors which have extremely short evolution timescales. So, their age estimates do not appear to be biased. Fig. \ref{slikica_podatki} shows no clear connection between mass and age of the BH. The objects with ages estimated from studies of their evolution (blue symbols) seem to be younger and somewhat less massive than the ones with age estimates based on their metallicity (green symbols). This may be a consequence of the Malmquist bias, which favours detailed studies of nearby objects. In contrast, only objects with large luminosities and thus containing a very massive BH are detected at large distances.

\section{Methodology}\label{metode}

The main novelty of our approach is to use age data to determine the mass of the BHs when they were formed. As noted above, masses and ages are subject to substantial uncertainties.

The top two panels in Fig. \ref{porazdelitev mas} plot the current distribution of masses for young (a) and old (b) objects. The shaded area is the combined PDF of the masses. Horizontal errorbars mark positions at 16\%, 50\%\ and 84\%\ of the combined PDF. Coloured lines plot the distributions for individual objects. Detached systems are plotted in red, X-ray binaries with ages determined from evolutionary studies are in blue, and those with age judged from metallicity or mass transfer rate are in green. Vertical bars plot the current most probable masses for individual BHs.

In further analysis,  probability distribution functions (PDF) were assumed to follow (asymmetric) Gaussian profiles. From this distribution, we randomly chose 100,000 values for masses and ages. Understandably, the ages were limited to positive values; moreover, we assumed that any BH formed at least 0.5~Gyr after the Big Bang. Using present mass and age, we were able to obtain the distribution of mass of the BH at its formation for each object as follows. Firstly, the metallicities of companion stars were converted into their ages by observational data in figure 6 by \citet{thompson18}. As described above this is assumed to be also the age of BH. The time when the BH was formed can be easily obtained by $t_0=t_U-t$, where $t_U$ is the age of the universe and $t$ age of BH. It can be connected to the scale factor at the formation of black hole $a_0$ in equation~\ref{potencni} by the polynomial approximation in \cite{salcido18} and the Planck values of cosmological parameters \citep{planck18}. Finally, we used relation~\ref{potencni} to determine $m_0$ assuming certain $n$. Repeating it for every previously chosen combination of mass and age we obtained the distribution of initial masses of BHs.

\section{Results}\label{rezultati}
We took two approaches. In first one, we checked how many initial masses fall under the TOV limit of $2.2$~M$_\odot$ for a specific value of $n$. In the second approach, we compared the distribution of initial masses of BH for certain $n$ with the distribution of masses of the youngest BHs. In both approaches, we assumed that a neutron star cannot be transformed into a black hole, either by collision of two neutron stars or by mass growth of a neutron star.

\subsection{Forming BHs with masses of neutron stars}

\begin{figure}
\centering
\begin{subfigure}[b]{\columnwidth}
  \includegraphics[width=\linewidth]{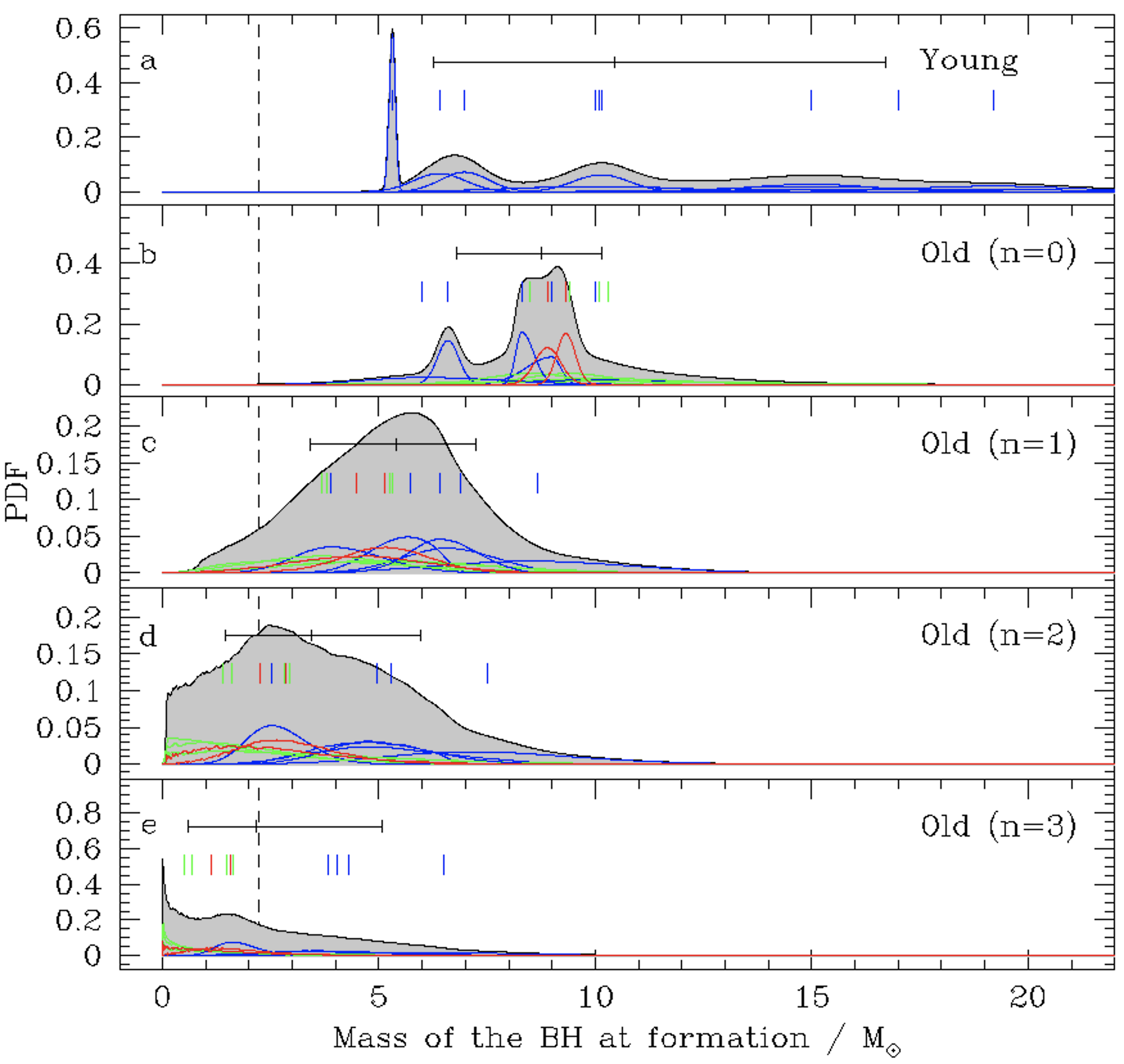}
\end{subfigure}
\caption{Probability distribution function of initial masses of BHs: (a) for BHs younger than 0.5~Gyr (data from table \ref{mlade}), (b) for BHs older than 0.5~Gyr if their mass does not change with time (data from table \ref{stare}), (c-e) respective distributions for older BHs if their mass scales with the indicated value of the power-law ($n$) of the scaling factor (eq.~\ref{potencni}). Coloured vertical lines indicate scaled modes of masses of individual BHs, and coloured lines are their mass distributions, following the colour scheme from Fig. \ref{slikica_podatki}. The horizontal bar marks the values at 16 per cent, 50 per cent, and 84 per cent of the combined mass distribution. Vertical dashed line is the mass of the TOV limit.}
\label{porazdelitev mas}
\end{figure}

The present-day mass distribution of BHs which are older than 0.5~Gyr (Fig. \ref{porazdelitev mas}b) is different from the corresponding distribution at their formation if BHs change their mass with the scale factor to the power of $n$. Figs. \ref{porazdelitev mas}c-e plot the results for $n=$1, 2, and 3, respectively. 

For $n>0$, the masses of BHs at their formation were smaller than the ones currently observed. The effect is moderate for $n=1$ but strong and very strong for $n=2$ and $n=3$, respectively. Smaller masses are not problematic per se, as long as the initial mass is not smaller than the Tolman–Oppenheimer–Volkoff (TOV) limit which is thought to mark the transition of compact objects from neutron stars to black holes. Its adopted value of 2.2~M$_\odot$ is plotted in Fig. \ref{porazdelitev mas} with a vertical dashed line.

As suggested already by \citet{rodriguez23} and  \citet{andrae23}, some BHs appear to have formed under the TOV limit if we assume $n=3$. Here we extend this approach to a larger sample of BHs and check what happens for other values of $n$. 

Fig. \ref{porazdelitev mas} shows that increasing $n$ moves the combined PDFs to substantially lower masses. The second column in Table \ref{nad TOV} shows that for $n=0$ nearly all objects have masses above the TOV limit. This is understandable, as $n=0$ implies no cosmic evolution of masses, and the objects we consider are very likely to contain compact objects more massive than 2.2~M$_\odot$. As we approach $n=3$, the fraction of objects which would be considered to contain a BH at the time of compact object formation falls to about $50$ per cent. This implies that neutron stars turning to BHs should have been common. Assuming that such a transition does not occur at all we can derive an even more stringent limit on $n$. The third column in Table \ref{nad TOV} shows the probability that every compact object in Table \ref{stare} formed with a mass above the TOV limit. Note that the probabilities for $n=2$ and $n=3$ are very small. Formally, $n=2$ is excluded at a $2.2\, \sigma$ and $n=3$ at a $4.0\,\sigma$ level. 

As an illustration, we quote probabilities for one specific X-ray binary in the last column of Table \ref{nad TOV}. MAXI J1659--152 has been subject to detailed evolutionary modelling, as has been explained in Sec.\ \ref{stare}, and a relatively small mass of its BH is rather well established (Table \ref{stare} and Appendix \ref{A2}). Table \ref{nad TOV} shows that there is just an 18 per cent probability that its compact object formed above the TOV limit if $n=3$ applies. 

\begin{table}
    \caption{Probabilities that the masses of BHs at formation are above the TOV limit of 2.2~M$_\odot$ for different $n$. The second column lists the fraction of the combined PDF of all objects, and the third column shows the probability that every BH from table \ref{stare} is above the limit. The last column lists probabilities for one well-studied object.}
    \centering
    \begin{tabular}{ cccc }
     \hline
n & fraction of BHs & every BH & MAXI J1659--152 \\
    \hline
0 & 0.9989 & 0.98761  & 0.9986 \\
1 & 0.94855 & 0.54461  & 0.9806 \\
2 & 0.7235 & 0.01442  & 0.7544 \\
3 & 0.4969 & 0.00003  & 0.1848 \\
    \hline
    \end{tabular}
    \label{nad TOV}
\end{table}

\subsection{Comparing mass distributions of young and old BHs}

Table \ref{mlade} lists ten BHs which are young, so their mass could not have changed yet, even if $n>0$. On the other hand, such values of $n$ may cause significant changes in mass distribution, as demonstrated by grey lines in Fig. \ref{slikica_podatki}. Here we assume that the distribution of BH masses at their formation does not change with time. So, we compare the sample of old  BHs to the young ones and study which value of $n$ best matches their masses at formation. Such an approach has limitations, as we consider only about a dozen objects in each group. Also, the two samples may be biased, as discussed below. Nevertheless, a rather steep dependence of masses on age for different values of $n$ (Fig. \ref{slikica_podatki}) suggests that a robust estimate of the value of $n$ can be obtained.

\subsubsection{Choice of comparison test}

The mass distribution of BHs is far from Gaussian (see Fig.~\ref{porazdelitev mas}), so any statistical tests assuming an approximate Gaussian distribution (like the Z-test or the Student T test) are not appropriate. Similarly, one should avoid the Kolmogorov-Smirnov, Kuiper's, or Wilcoxon-Mann-Whitney test because they count the number of data below some specific value (see however section~\ref{diskusija} for a comparison of results to the Kolmogorov-Smirnov test). This can be a problem, as ranking may be insensitive to changes in the tails of the distribution of a rather small number of objects.

Having this in mind, we used the Anderson-Darling test. It considers these problems and puts more weight on tails, where the difference between mass distributions is the largest.

\subsubsection{Anderson-Darling test}
\label{sec_anderson}
We began our analysis by randomly choosing a value from (asymmetric) Gaussian distribution for mass and age, as given in Table \ref{stare}. Then, we used those randomly chosen values to calculate the distribution of masses of BHs at their formation for a given $n$. The results were compared to a randomly sampled mass distribution of young BHs from Table~\ref{mlade}. Similarity was evaluated by calculating the value of the statistics of non-parametric k-sample ($k=2$) Anderson-Darling test:
\begin{equation}
    A^2=\frac{1}{N}\sum_{i=1}^{2} \frac{1}{n_i} \sum_{j=1}^{N-1} \frac{(NM_{ij}-jn_i)^2}{j(N-j)},
\end{equation}
where $n_i$ is the number of data in $i$-th sample and $N=n_1+n_2$. $Z_1<...<Z_N$ is the ordered sample of masses from both samples and $M_{ij}$ number of masses in the i-th sample not greater than $Z_j$. 

\begin{figure}
\centering
\begin{subfigure}[b]{\columnwidth}
  \includegraphics[width=\linewidth]{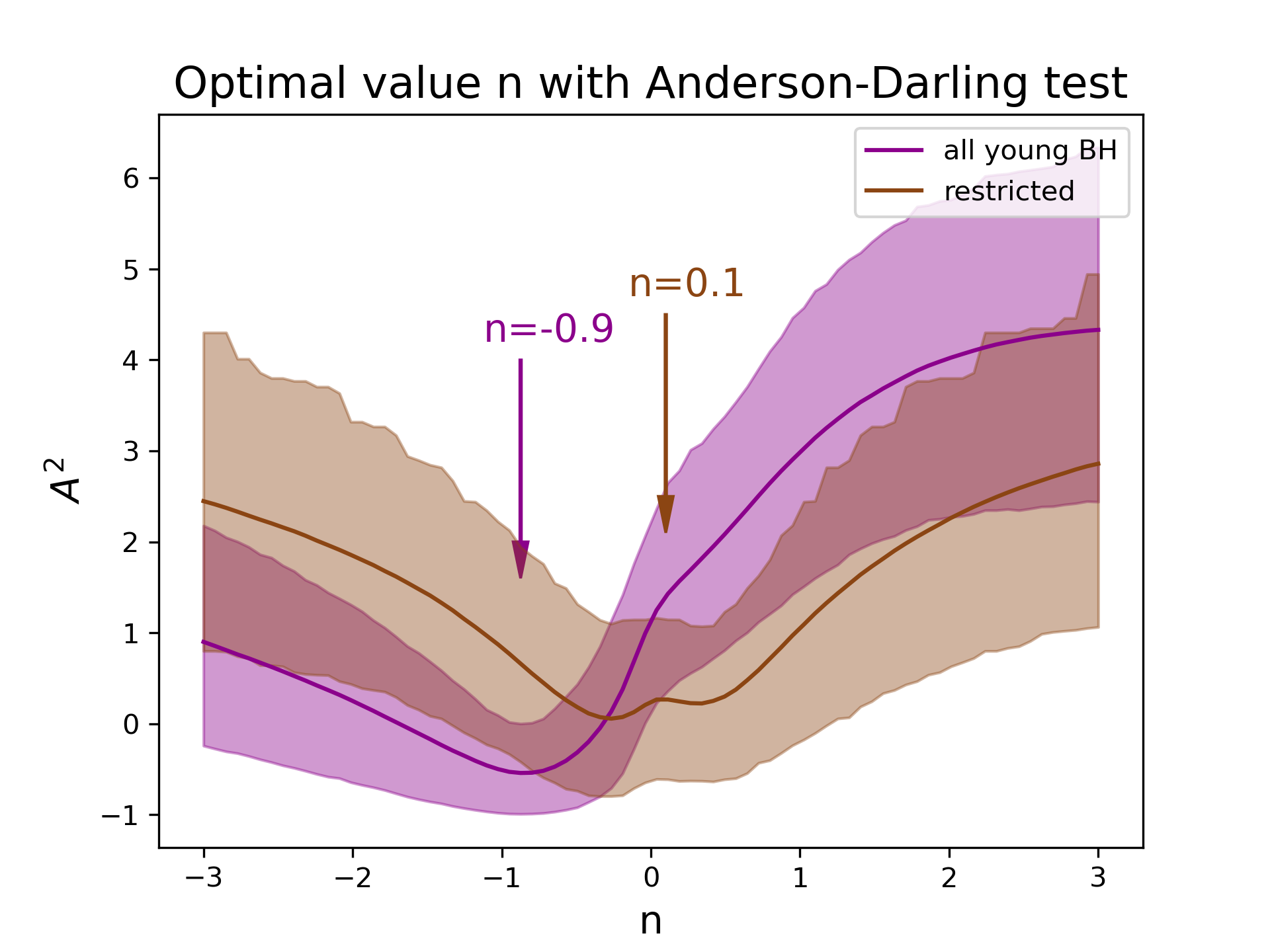}
\end{subfigure}
\caption{Similarity of mass distributions of the youngest and the older BHs at their formation. We plot the value of Anderson-Darling statistic $A^2$ for different values of $n$. Purple lines compare the whole samples, and brown exclude any young BHs from Table \ref{mlade}. These were judged to be young because their companion is massive and therefore young. Shaded areas denote the confidence level of 90 per cent.
 }
\label{anderson}
\end{figure}

The result as a function of $n$ for entire young and old samples is shown as a purple line in Fig.~\ref{anderson}. The optimal value for n is $n = -0.9^{+1.3}_{-4.6}$ (1$\sigma$). We can exclude $n=3$ with a confidence level of 87 per cent.

One may wonder why the result favours moderately negative values of $n$. Comparison of Fig. \ref{porazdelitev mas}a and b shows that some of the young black holes are very massive. We believe this reflects a bias, because the age of many young BHs was established by their massive and therefore young companion. If a companion is massive, a BH is more likely to be massive too. This lowers the derived value of $n$. 

In order to estimate the amount of this bias, we can do the following test. As a rather extreme choice, we exclude all of the young systems with their age determined by a young companion. Because such a choice leaves us with only three young objects, we opt to count two of the objects (namely GS 1354--64 and A0620--00) from the old group as young. This is a reasonable approximation since their ages are $2\pm 1$ Gyr and $2\pm 2$ Gyr, while the average of other ages in the old group is $7$ Gyr. The result is a brown line in Fig.~\ref{anderson}, with an optimal value of $n=0.1^{+2.0}_{-3.1}$ (1$\sigma$). As expected, this is larger than that derived for the whole samples. Although this result is of limited reliability because of a small number of data, it suggests that our value for the whole sample was a bit underestimated. We can conclude that the true value of $n$ equals zero within errorbars and cannot be much larger than $0.5$.

Negative values of $n$ do not have any theoretical explanation. Evaporation of BHs via Hawking radiation could work in principle but on entirely different mass range. We conclude that $n$ should be very close to 0, and any deviations from this can be explained by small number statistics and biases similar to the one mentioned above. 

\section{Discussion}\label{diskusija}

We have shown that for $n \gtrsim 2$, many of the BHs would have to be formed with masses below the TOV limit (Table \ref{nad TOV}). Even at $n=1$ there is just a 50 per cent chance that all of the BHs older than 0.5~Gyr have been formed with masses above this limit. But note that the last statement is sensitive to observational errors because we estimate the property to be fulfilled by all objects, so an unexpected observational error for any of them could change the result. We conclude that the TOV favours $n \leq 1$. 

One may argue that the TOV limit is not important because BHs crossing it could result from a stellar merge. We find this improbable since the orbit needs quite a while to tighten enough. Furthermore, it would mean that we have a very tight system of at least three stars (of which two neutron stars formed by former supernova explosions) at the beginning, which is highly unlikely. 

In the second approach, we compared the distributions of masses of young and old objects, again deriving that $n \lesssim 1$, with $n=0$ the preferred solution. We used the Anderson-Darling test to derive these numbers in Sec.\ \ref{sec_anderson}. For completeness, we rerun the same analysis but with a Kolmogorov-Smirnov test. The result  $n=-0.7^{+1.1}_{-3.9}$ (1$\sigma$) is almost identical, so the choice of statistical test does not influence the results substantially.

The smallest mass of the BHs among young objects is $5.31 \pm 0.07$~M$_\odot$ (in GRO J1655--40), and the corresponding lightest BH among old objects has $6^{+1.8}_{-1.3}$~M$_\odot$ (in MAXI J1659--152). So, there is a scarcity of BHs with masses just above the TOV limit. This is understandable because observational errors would force such objects into an uncertain category in the sense that they may contain a neutron star or a BH. Such objects may influence the derived value of $n$. If such objects were old they would force an even more stringent limit on $n$ to avoid TOV limit crossings. But if they were young, typical masses of young objects would be smaller, allowing for larger values of $n$ when comparing young and old BH populations. 

For example, consider SN~1997D, which may contain a BH or a neutron star \citep{zampieri98,benetti01}. It is a young object with parameters listed at the end of Table \ref{mlade}. We ran the Anderson-Darling test with an addition of this object. The corresponding $n = -0.8^{+1.9}_{-4.3}$ is very similar to our results from  section~\ref{rezultati}. Such low mass BHs are only important if  they are very common, which does not seem to be the case \citep{belczynski12}. 

\subsection{The role of accretion}
Until now we ignored the mass transferred via accretion. We note that accretion can only increase the mass of the BH with time, so it may increase the derived value of $n$. As our values of $n$ are close to zero, we conclude that accretion does not seem significant. Also, the influence of accretion has the wrong sign to be able to justify the cosmic scaling of BH masses.

Nevertheless, most of our objects are X-ray binaries, where accretion is present. So it is instructive to make a simple estimate of an upper limit of its influence. The average mass of BHs in table~\ref{ostale} is $8.8$~ M$_\odot$ and, if we convert their average age to a scale factor at their formation, we get $a \approx 0.66$. Orbital periods in Table~\ref{ostale} indicate that MAXI J1659--152 and XTE J1118+480 are the tightest systems, so their mass transfer should be most significant. \citet{kuulkers13} and \citet{gualandris05} estimate that the companion lost less than $1.3$~M$_\odot$ and about $1.2$~M$_\odot$ between formation and the present in these two systems, respectively. These correspond to an upper limit of mass growth of these two BHs. Combination of these numbers with the ages of these two systems yields an average accretion rate over the system lifetime of about $3 \cdot 10^{-9}$~M$_\odot$~yr$^{-1}$ in both cases. Combining these numbers with equation~\ref{potencni} shows that accretion may increase the effective value of $n$ by $\Delta n < 0.3$. Note that this is an upper limit, but even at this value it is within the errors of our results.

\subsection{Comparison with other authors}\label{primerjava}
Results of our first approach are  consistent with results of \citet{rodriguez23} and \citet{andrae23}. With more data, we get even more restrictive conditions. Our results are also compatible with LIGO-Virgo-KAGRA \citep{amendola23}, since they found strong restrictions on $n>2.4$. For $n=0.5$ they found no relevant constraints, which is also consistent with our work.

The results for supermassive BHs appear  different from stellar-mass BHs since the scaling power $n\approx 3$ found by \citet{farrah23b} is inconsistent with our results. On the other hand, results for supermassive BHs at large redshifts \citep{lei23,cadoni23} appear  consistent with ours.

Our results for stellar-mass BHs are  therefore inconsistent with predictions for GEODE by \citet{croker19}. Similar conclusions can be drawn for the power law dependence claimed by \citet{cadoni23}, although with a lower confidence level.

\section{Conclusions}

We presented a new approach to study the cosmological growth of black holes. It uses X-ray binaries and their properties from the literature to estimate the ages of BHs and, therefore, their mass growth rate.

We used primarily the masses and the ages of BHs. BHs were divided into two age groups. Young BHs are those with ages less than $0.5$~Gyr, with their youth supported by four types of arguments. Such BHs could not grow significantly in their lifetime. The old group consists of objects with larger ages, estimated from the evolutionary history of the binary system and, in some cases, from metallicity of the companion star. 

Two different approaches were used to study which power $n$ is consistent with the supposed cosmological growth of BHs in our observational sample. The first approach was an extension of methods by \citet{rodriguez23} and \citet{andrae23} where we estimated the fraction of BHs that were born with an initial mass larger than the Tolman–Oppenheimer–Volkof limit of 2.2~M$_\odot$. We've shown that half of the BHs would have to be born with masses below this limit if $n=3$. Also, there is a formal probability of just $3 \cdot 10^{-5}$ that {\it all} of the BHs were born with mass above this limit. Though the last statement has a formal significance of $4\,\sigma$ we note that it is sensitive to errors on individual objects. Nevertheless, this approach favours $n \leq 1$.  

As the second approach, we use age estimates of BHs older than 0.5~Gyr.  For these, we calculated the distribution of initial masses and then compared it using the Anderson-Darling statistic test with the distribution of masses of young BHs. Here, we crucially assumed that the distribution of initial masses of BHs did not change over cosmic history. This enabled us to estimate the power of $n = -0.9^{+1.3}_{-4.6}$. We can also exclude $n=3$ with a confidence level of 87 per cent, which means that our results are not consistent with the results for supermassive black holes by \citet{farrah23b}. If we correct for the bias that masses of BHs and their companions are correlated, we get the value $n\approx 0$. In other words, our results show that any cosmological growth of masses of stellar BHs is unlikely, with values $n\geq 2$ excluded at a greater than $2\,\sigma$ level.

Our procedure can be easily applied to more data. Mass estimates for the compact object are available for many X-ray binaries, but their ages are missing. A helpful clue, which we exploited in this work, is the nature of the companion star, which may constrain the past evolution of the system and, therefore, its age. On the other hand, the search for supernovae and their remnants can substantially increase the number of known young BH binaries. Another possibility is to look for more detached BH binaries, which are being discovered by the Gaia mission. Their accurate spectral and orbital parameters and the absence of biases which complicate the interpretation of compact systems makes these discoveries very special and hugely valuable.

\section*{Acknowledgements}
We thank the reviewer Christopher Tout for very useful comments that helped us improve the clarity and presentation of our work.
TZ acknowledges financial support of the Slovenian Research Agency
(research core funding No. P1-0188) and the European Space Agency
(Prodex Experiment Arrangement No. 4000142234).

\section*{Data availability}

This paper uses exclusively data from the literature. All data and their
sources are listed in Tables 1 and 2.

\bibliographystyle{mnras}
\bibliography{main}

\begin{thebibliography}{}
\makeatletter
\relax
\def\mn@urlcharsother{\let\do\@makeother \do\$\do\&\do\#\do\^\do\_\do\%\do\~}
\def\mn@doi{\begingroup\mn@urlcharsother \@ifnextchar [ {\mn@doi@} {\mn@doi@[]}}
\def\mn@doi@[#1]#2{\def\@tempa{#1}\ifx\@tempa\@empty \href {http://dx.doi.org/#2} {doi:#2}\else \href {http://dx.doi.org/#2} {#1}\fi \endgroup}
\def\mn@eprint#1#2{\mn@eprint@#1:#2::\@nil}
\def\mn@eprint@arXiv#1{\href {http://arxiv.org/abs/#1} {{\tt arXiv:#1}}}
\def\mn@eprint@dblp#1{\href {http://dblp.uni-trier.de/rec/bibtex/#1.xml} {dblp:#1}}
\def\mn@eprint@#1:#2:#3:#4\@nil{\def\@tempa {#1}\def\@tempb {#2}\def\@tempc {#3}\ifx \@tempc \@empty \let \@tempc \@tempb \let \@tempb \@tempa \fi \ifx \@tempb \@empty \def\@tempb {arXiv}\fi \@ifundefined {mn@eprint@\@tempb}{\@tempb:\@tempc}{\expandafter \expandafter \csname mn@eprint@\@tempb\endcsname \expandafter{\@tempc}}}

\bibitem[\protect\citeauthoryear{{Abubekerov}, {Antokhina}, {Gostev}, {Cherepashchuk}  \& {Shimansky}}{{Abubekerov} et~al.}{2016}]{abubekerov16}
{Abubekerov} M.~K.,  {Antokhina} E.~A.,  {Gostev} N.~Y.,  {Cherepashchuk} A.~M.,   {Shimansky} V.~V.,  2016, \mn@doi [Astronomy Reports] {10.1134/S1063772916120015}, \href {https://ui.adsabs.harvard.edu/abs/2016ARep...60.1029A} {60, 1029}

\bibitem[\protect\citeauthoryear{{Amendola}, {Rodrigues}, {Kumar}  \& {Quartin}}{{Amendola} et~al.}{2023}]{amendola23}
{Amendola} L.,  {Rodrigues} D.~C.,  {Kumar} S.,   {Quartin} M.,  2023, \mn@doi [arXiv e-prints] {10.48550/arXiv.2307.02474}, \href {https://ui.adsabs.harvard.edu/abs/2023arXiv230702474A} {p. arXiv:2307.02474}

\bibitem[\protect\citeauthoryear{{Anders} et~al.,}{{Anders} et~al.}{2022}]{anders22}
{Anders} F.,  et~al., 2022, \mn@doi [\aap] {10.1051/0004-6361/202142369}, \href {https://ui.adsabs.harvard.edu/abs/2022A&A...658A..91A} {658, A91}

\bibitem[\protect\citeauthoryear{{Andrae} \& {El-Badry}}{{Andrae} \& {El-Badry}}{2023}]{andrae23}
{Andrae} R.,  {El-Badry} K.,  2023, \mn@doi [\aap] {10.1051/0004-6361/202346350}, \href {https://ui.adsabs.harvard.edu/abs/2023A&A...673L..10A} {673, L10}

\bibitem[\protect\citeauthoryear{{Atri} et~al.,}{{Atri} et~al.}{2020}]{atri20}
{Atri} P.,  et~al., 2020, \mn@doi [\mnras] {10.1093/mnrasl/slaa010}, \href {https://ui.adsabs.harvard.edu/abs/2020MNRAS.493L..81A} {493, L81}

\bibitem[\protect\citeauthoryear{{Bartolomeo Koninckx}, {De Vito}  \& {Benvenuto}}{{Bartolomeo Koninckx} et~al.}{2023}]{bartolomeo23}
{Bartolomeo Koninckx} L.,  {De Vito} M.~A.,   {Benvenuto} O.~G.,  2023, \mn@doi [\aap] {10.1051/0004-6361/202346571}, \href {https://ui.adsabs.harvard.edu/abs/2023A&A...674A..97B} {674, A97}

\bibitem[\protect\citeauthoryear{{Beer} \& {Podsiadlowski}}{{Beer} \& {Podsiadlowski}}{2002}]{beer02}
{Beer} M.~E.,  {Podsiadlowski} P.,  2002, \mn@doi [\mnras] {10.1046/j.1365-8711.2002.05189.x}, \href {https://ui.adsabs.harvard.edu/abs/2002MNRAS.331..351B} {331, 351}

\bibitem[\protect\citeauthoryear{{Belczynski}, {Wiktorowicz}, {Fryer}, {Holz}  \& {Kalogera}}{{Belczynski} et~al.}{2012}]{belczynski12}
{Belczynski} K.,  {Wiktorowicz} G.,  {Fryer} C.~L.,  {Holz} D.~E.,   {Kalogera} V.,  2012, \mn@doi [\apj] {10.1088/0004-637X/757/1/91}, \href {https://ui.adsabs.harvard.edu/abs/2012ApJ...757...91B} {757, 91}

\bibitem[\protect\citeauthoryear{{Benetti} et~al.,}{{Benetti} et~al.}{2001}]{benetti01}
{Benetti} S.,  et~al., 2001, \mn@doi [\mnras] {10.1046/j.1365-8711.2001.04122.x}, \href {https://ui.adsabs.harvard.edu/abs/2001MNRAS.322..361B} {322, 361}

\bibitem[\protect\citeauthoryear{{Bezayiff}}{{Bezayiff}}{2006}]{bezayiff06}
{Bezayiff} N.,  2006, PhD thesis, University of California, Santa Cruz

\bibitem[\protect\citeauthoryear{{Bhargava}, {Belloni}, {Bhattacharya}  \& {Misra}}{{Bhargava} et~al.}{2019}]{bhargava19}
{Bhargava} Y.,  {Belloni} T.,  {Bhattacharya} D.,   {Misra} R.,  2019, \mn@doi [\mnras] {10.1093/mnras/stz1774}, \href {https://ui.adsabs.harvard.edu/abs/2019MNRAS.488..720B} {488, 720}

\bibitem[\protect\citeauthoryear{{Bhuvana G.}, {Radhika}, {Mandal}, {Nandi}  \& {Agrawal}}{{Bhuvana G.} et~al.}{2021}]{bhuvana21}
{Bhuvana G.} R.,  {Radhika} D.,  {Mandal} S.,  {Nandi} A.,   {Agrawal} V.~K.,  2021, in 43rd COSPAR Scientific Assembly. Held 28 January - 4 February. p.~1588

\bibitem[\protect\citeauthoryear{{Binder}, {Eracleous}, {Plucinsky}  \& {Williams}}{{Binder} et~al.}{2019}]{binder19}
{Binder} B.,  {Eracleous} M.,  {Plucinsky} P.~P.,   {Williams} B.~F.,  2019, {What is the Mass of the NGC 300 X-1 Black Hole?}, HST Proposal. Cycle 27, ID. \#15999

\bibitem[\protect\citeauthoryear{{Binder} et~al.,}{{Binder} et~al.}{2021}]{binder21}
{Binder} B.~A.,  et~al., 2021, \mn@doi [\apj] {10.3847/1538-4357/abe6a9}, \href {https://ui.adsabs.harvard.edu/abs/2021ApJ...910...74B} {910, 74}

\bibitem[\protect\citeauthoryear{{Blundell}, {Bowler}  \& {Schmidtobreick}}{{Blundell} et~al.}{2008}]{blundell08}
{Blundell} K.~M.,  {Bowler} M.~G.,   {Schmidtobreick} L.,  2008, \mn@doi [\apjl] {10.1086/588027}, \href {https://ui.adsabs.harvard.edu/abs/2008ApJ...678L..47B} {678, L47}

\bibitem[\protect\citeauthoryear{{Bowler}}{{Bowler}}{2010}]{bowler10}
{Bowler} M.~G.,  2010, \mn@doi [arXiv e-prints] {10.48550/arXiv.1006.5213}, \href {https://ui.adsabs.harvard.edu/abs/2010arXiv1006.5213B} {p. arXiv:1006.5213}

\bibitem[\protect\citeauthoryear{{Bowler}}{{Bowler}}{2018}]{bowler18}
{Bowler} M.~G.,  2018, \mn@doi [\aap] {10.1051/0004-6361/201834121}, \href {https://ui.adsabs.harvard.edu/abs/2018A&A...619L...4B} {619, L4}

\bibitem[\protect\citeauthoryear{{Boyd}, {Smale}  \& {Dolan}}{{Boyd} et~al.}{2001}]{boyd01}
{Boyd} P.~T.,  {Smale} A.~P.,   {Dolan} J.~F.,  2001, \mn@doi [\apj] {10.1086/321511}, \href {https://ui.adsabs.harvard.edu/abs/2001ApJ...555..822B} {555, 822}

\bibitem[\protect\citeauthoryear{{Brocksopp}, {Tarasov}, {Lyuty}  \& {Roche}}{{Brocksopp} et~al.}{1999}]{brocksopp99}
{Brocksopp} C.,  {Tarasov} A.~E.,  {Lyuty} V.~M.,   {Roche} P.,  1999, \mn@doi [\aap] {10.48550/arXiv.astro-ph/9812077}, \href {https://ui.adsabs.harvard.edu/abs/1999A&A...343..861B} {343, 861}

\bibitem[\protect\citeauthoryear{{Cadoni}, {Sanna}, {Pitzalis}, {Banerjee}, {Murgia}, {Hazra}  \& {Branchesi}}{{Cadoni} et~al.}{2023}]{cadoni23}
{Cadoni} M.,  {Sanna} A.~P.,  {Pitzalis} M.,  {Banerjee} B.,  {Murgia} R.,  {Hazra} N.,   {Branchesi} M.,  2023, \mn@doi [arXiv e-prints] {10.48550/arXiv.2306.11588}, \href {https://ui.adsabs.harvard.edu/abs/2023arXiv230611588C} {p. arXiv:2306.11588}

\bibitem[\protect\citeauthoryear{{Cantrell} et~al.,}{{Cantrell} et~al.}{2010}]{cantrell10}
{Cantrell} A.~G.,  et~al., 2010, \mn@doi [\apj] {10.1088/0004-637X/710/2/1127}, \href {https://ui.adsabs.harvard.edu/abs/2010ApJ...710.1127C} {710, 1127}

\bibitem[\protect\citeauthoryear{{Carpano}, {Pollock}, {Prestwich}, {Crowther}, {Wilms}, {Yungelson}  \& {Ehle}}{{Carpano} et~al.}{2007}]{carpano07}
{Carpano} S.,  {Pollock} A.~M.~T.,  {Prestwich} A.,  {Crowther} P.,  {Wilms} J.,  {Yungelson} L.,   {Ehle} M.,  2007, \mn@doi [\aap] {10.1051/0004-6361:20077363}, \href {https://ui.adsabs.harvard.edu/abs/2007A&A...466L..17C} {466, L17}

\bibitem[\protect\citeauthoryear{{Casares}, {Zurita}, {Shahbaz}, {Charles}  \& {Fender}}{{Casares} et~al.}{2004}]{casares04}
{Casares} J.,  {Zurita} C.,  {Shahbaz} T.,  {Charles} P.~A.,   {Fender} R.~P.,  2004, \mn@doi [\apjl] {10.1086/425145}, \href {https://ui.adsabs.harvard.edu/abs/2004ApJ...613L.133C} {613, L133}

\bibitem[\protect\citeauthoryear{{Casares} et~al.,}{{Casares} et~al.}{2009}]{casares09}
{Casares} J.,  et~al., 2009, \mn@doi [\apjs] {10.1088/0067-0049/181/1/238}, \href {https://ui.adsabs.harvard.edu/abs/2009ApJS..181..238C} {181, 238}

\bibitem[\protect\citeauthoryear{{Casares}, {Jonker}  \& {Israelian}}{{Casares} et~al.}{2017}]{casares17}
{Casares} J.,  {Jonker} P.~G.,   {Israelian} G.,  2017, in {Alsabti} A.~W.,  {Murdin} P.,  eds, , Handbook of Supernovae.
Springer, p.~1499, \mn@doi{10.1007/978-3-319-21846-5_111}

\bibitem[\protect\citeauthoryear{{Casares}, {Mu{\~n}oz-Darias}, {Mata S{\'a}nchez}, {Charles}, {Torres}, {Armas Padilla}, {Fender}  \& {Garc{\'\i}a-Rojas}}{{Casares} et~al.}{2019}]{casares19}
{Casares} J.,  {Mu{\~n}oz-Darias} T.,  {Mata S{\'a}nchez} D.,  {Charles} P.~A.,  {Torres} M.~A.~P.,  {Armas Padilla} M.,  {Fender} R.~P.,   {Garc{\'\i}a-Rojas} J.,  2019, \mn@doi [\mnras] {10.1093/mnras/stz1793}, \href {https://ui.adsabs.harvard.edu/abs/2019MNRAS.488.1356C} {488, 1356}

\bibitem[\protect\citeauthoryear{{Chakrabarti} et~al.,}{{Chakrabarti} et~al.}{2023}]{chakrabarti23}
{Chakrabarti} S.,  et~al., 2023, \mn@doi [\aj] {10.3847/1538-3881/accf21}, \href {https://ui.adsabs.harvard.edu/abs/2023AJ....166....6C} {166, 6}

\bibitem[\protect\citeauthoryear{{Chakraborty}, {Navale}, {Ratheesh}  \& {Bhattacharyya}}{{Chakraborty} et~al.}{2020}]{chakraborty20}
{Chakraborty} S.,  {Navale} N.,  {Ratheesh} A.,   {Bhattacharyya} S.,  2020, \mn@doi [\mnras] {10.1093/mnras/staa2711}, \href {https://ui.adsabs.harvard.edu/abs/2020MNRAS.498.5873C} {498, 5873}

\bibitem[\protect\citeauthoryear{{Chand}, {Dewangan}, {Thakur}, {Tripathi}  \& {Agrawal}}{{Chand} et~al.}{2022}]{chand22}
{Chand} S.,  {Dewangan} G.~C.,  {Thakur} P.,  {Tripathi} P.,   {Agrawal} V.~K.,  2022, \mn@doi [\apj] {10.3847/1538-4357/ac7154}, \href {https://ui.adsabs.harvard.edu/abs/2022ApJ...933...69C} {933, 69}

\bibitem[\protect\citeauthoryear{{Chatterjee}, {Debnath}, {Jana}  \& {Chakrabarti}}{{Chatterjee} et~al.}{2019}]{chatterjee19}
{Chatterjee} D.,  {Debnath} D.,  {Jana} A.,   {Chakrabarti} S.~K.,  2019, \mn@doi [\apss] {10.1007/s10509-019-3495-2}, \href {https://ui.adsabs.harvard.edu/abs/2019Ap&SS.364...14C} {364, 14}

\bibitem[\protect\citeauthoryear{{Chauhan} et~al.,}{{Chauhan} et~al.}{2019}]{Chauhan19}
{Chauhan} J.,  et~al., 2019, \mn@doi [\mnras] {10.1093/mnrasl/slz113}, \href {https://ui.adsabs.harvard.edu/abs/2019MNRAS.488L.129C} {488, L129}

\bibitem[\protect\citeauthoryear{{Chen}}{{Chen}}{2011}]{chen11}
{Chen} T.,  2011, in {Romero} G.~E.,  {Sunyaev} R.~A.,   {Belloni} T.,  eds,  IAU Symposium Vol. 275, Jets at All Scales. pp 327--328, \mn@doi{10.1017/S1743921310016339}

\bibitem[\protect\citeauthoryear{{Cherepashchuk}, {Postnov}  \& {Belinski}}{{Cherepashchuk} et~al.}{2019}]{chere19}
{Cherepashchuk} A.~M.,  {Postnov} K.~A.,   {Belinski} A.~A.,  2019, \mn@doi [\mnras] {10.1093/mnras/stz610}, \href {https://ui.adsabs.harvard.edu/abs/2019MNRAS.485.2638C} {485, 2638}

\bibitem[\protect\citeauthoryear{{Cherepashchuk}, {Belinski}, {Dodin}  \& {Postnov}}{{Cherepashchuk} et~al.}{2021}]{chere21}
{Cherepashchuk} A.~M.,  {Belinski} A.~A.,  {Dodin} A.~V.,   {Postnov} K.~A.,  2021, \mn@doi [\mnras] {10.1093/mnrasl/slab083}, \href {https://ui.adsabs.harvard.edu/abs/2021MNRAS.507L..19C} {507, L19}

\bibitem[\protect\citeauthoryear{{Combi}, {Bronfman}  \& {Mirabel}}{{Combi} et~al.}{2007}]{combi07}
{Combi} J.~A.,  {Bronfman} L.,   {Mirabel} I.~F.,  2007, \mn@doi [\aap] {10.1051/0004-6361:20077156}, \href {https://ui.adsabs.harvard.edu/abs/2007A&A...467..597C} {467, 597}

\bibitem[\protect\citeauthoryear{{Cook}, {Patterson}, {Buczynski}  \& {Fried}}{{Cook} et~al.}{2000}]{cook00}
{Cook} L.,  {Patterson} J.,  {Buczynski} D.,   {Fried} R.,  2000, \iaucirc, \href {https://ui.adsabs.harvard.edu/abs/2000IAUC.7397....2C} {7397, 2}

\bibitem[\protect\citeauthoryear{{Croker} \& {Weiner}}{{Croker} \& {Weiner}}{2019}]{croker19}
{Croker} K.~S.,  {Weiner} J.~L.,  2019, \mn@doi [\apj] {10.3847/1538-4357/ab32da}, \href {https://ui.adsabs.harvard.edu/abs/2019ApJ...882...19C} {882, 19}

\bibitem[\protect\citeauthoryear{{Crowther}, {Barnard}, {Carpano}, {Clark}, {Dhillon}  \& {Pollock}}{{Crowther} et~al.}{2010}]{crowther10}
{Crowther} P.~A.,  {Barnard} R.,  {Carpano} S.,  {Clark} J.~S.,  {Dhillon} V.~S.,   {Pollock} A.~M.~T.,  2010, \mn@doi [\mnras] {10.1111/j.1745-3933.2010.00811.x}, \href {https://ui.adsabs.harvard.edu/abs/2010MNRAS.403L..41C} {403, L41}

\bibitem[\protect\citeauthoryear{{Dong}, {Liu}, {Tuo}, {Steiner}, {Ge}, {Garc{\'\i}a}  \& {Cao}}{{Dong} et~al.}{2022}]{dong22}
{Dong} Y.,  {Liu} Z.,  {Tuo} Y.,  {Steiner} J.~F.,  {Ge} M.,  {Garc{\'\i}a} J.~A.,   {Cao} X.,  2022, \mn@doi [\mnras] {10.1093/mnras/stac1466}, \href {https://ui.adsabs.harvard.edu/abs/2022MNRAS.514.1422D} {514, 1422}

\bibitem[\protect\citeauthoryear{{El-Badry} et~al.,}{{El-Badry} et~al.}{2023a}]{el-badry23a}
{El-Badry} K.,  et~al., 2023a, \mn@doi [\mnras] {10.1093/mnras/stac3140}, \href {https://ui.adsabs.harvard.edu/abs/2023MNRAS.518.1057E} {518, 1057}

\bibitem[\protect\citeauthoryear{{El-Badry} et~al.,}{{El-Badry} et~al.}{2023b}]{el-badry23b}
{El-Badry} K.,  et~al., 2023b, \mn@doi [\mnras] {10.1093/mnras/stad799}, \href {https://ui.adsabs.harvard.edu/abs/2023MNRAS.521.4323E} {521, 4323}

\bibitem[\protect\citeauthoryear{{Esin}, {Kuulkers}, {McClintock}  \& {Narayan}}{{Esin} et~al.}{2000}]{esin00}
{Esin} A.~A.,  {Kuulkers} E.,  {McClintock} J.~E.,   {Narayan} R.,  2000, \mn@doi [\apj] {10.1086/308615}, \href {https://ui.adsabs.harvard.edu/abs/2000ApJ...532.1069E} {532, 1069}

\bibitem[\protect\citeauthoryear{{Farrah} et~al.,}{{Farrah} et~al.}{2023}]{farrah23b}
{Farrah} D.,  et~al., 2023, \mn@doi [\apjl] {10.3847/2041-8213/acb704}, \href {https://ui.adsabs.harvard.edu/abs/2023ApJ...944L..31F} {944, L31}

\bibitem[\protect\citeauthoryear{{Gelino}, {Balman}, {K{\i}z{\i}lo{\v{g}}lu}, {Y{\i}lmaz}, {Kalemci}  \& {Tomsick}}{{Gelino} et~al.}{2006}]{gelino06}
{Gelino} D.~M.,  {Balman} {\c{S}}.,  {K{\i}z{\i}lo{\v{g}}lu} {\"U}.,  {Y{\i}lmaz} A.,  {Kalemci} E.,   {Tomsick} J.~A.,  2006, \mn@doi [\apj] {10.1086/500924}, \href {https://ui.adsabs.harvard.edu/abs/2006ApJ...642..438G} {642, 438}

\bibitem[\protect\citeauthoryear{{Gonz{\'a}lez Hern{\'a}ndez}, {Rebolo}  \& {Israelian}}{{Gonz{\'a}lez Hern{\'a}ndez} et~al.}{2008}]{gonzalez08}
{Gonz{\'a}lez Hern{\'a}ndez} J.~I.,  {Rebolo} R.,   {Israelian} G.,  2008, \mn@doi [\aap] {10.1051/0004-6361:20077141}, \href {https://ui.adsabs.harvard.edu/abs/2008A&A...478..203G} {478, 203}

\bibitem[\protect\citeauthoryear{{Gonz{\'a}lez Hern{\'a}ndez}, {Casares}, {Rebolo}, {Israelian}, {Filippenko}  \& {Chornock}}{{Gonz{\'a}lez Hern{\'a}ndez} et~al.}{2011}]{gonzalez11}
{Gonz{\'a}lez Hern{\'a}ndez} J.~I.,  {Casares} J.,  {Rebolo} R.,  {Israelian} G.,  {Filippenko} A.~V.,   {Chornock} R.,  2011, \mn@doi [\apj] {10.1088/0004-637X/738/1/95}, \href {https://ui.adsabs.harvard.edu/abs/2011ApJ...738...95G} {738, 95}

\bibitem[\protect\citeauthoryear{{Gonz{\'a}lez Hern{\'a}ndez}, {Rebolo}  \& {Casares}}{{Gonz{\'a}lez Hern{\'a}ndez} et~al.}{2012}]{gonzalez12}
{Gonz{\'a}lez Hern{\'a}ndez} J.~I.,  {Rebolo} R.,   {Casares} J.,  2012, \mn@doi [\apjl] {10.1088/2041-8205/744/2/L25}, \href {https://ui.adsabs.harvard.edu/abs/2012ApJ...744L..25G} {744, L25}

\bibitem[\protect\citeauthoryear{{Goranskij}, {Barsukova}  \& {Burenkov}}{{Goranskij} et~al.}{2003}]{goranskij03}
{Goranskij} V.~P.,  {Barsukova} E.~A.,   {Burenkov} A.~N.,  2003, \mn@doi [Astronomy Reports] {10.1134/1.1611215}, \href {https://ui.adsabs.harvard.edu/abs/2003ARep...47..740G} {47, 740}

\bibitem[\protect\citeauthoryear{{Greene}, {Bailyn}  \& {Orosz}}{{Greene} et~al.}{2001}]{greene01}
{Greene} J.,  {Bailyn} C.~D.,   {Orosz} J.~A.,  2001, \mn@doi [\apj] {10.1086/321411}, \href {https://ui.adsabs.harvard.edu/abs/2001ApJ...554.1290G} {554, 1290}

\bibitem[\protect\citeauthoryear{{Gualandris}, {Colpi}, {Portegies Zwart}  \& {Possenti}}{{Gualandris} et~al.}{2005}]{gualandris05}
{Gualandris} A.,  {Colpi} M.,  {Portegies Zwart} S.,   {Possenti} A.,  2005, \mn@doi [\apj] {10.1086/426126}, \href {https://ui.adsabs.harvard.edu/abs/2005ApJ...618..845G} {618, 845}

\bibitem[\protect\citeauthoryear{{Heida}, {Jonker}, {Torres}  \& {Chiavassa}}{{Heida} et~al.}{2017}]{heida17}
{Heida} M.,  {Jonker} P.~G.,  {Torres} M.~A.~P.,   {Chiavassa} A.,  2017, \mn@doi [\apj] {10.3847/1538-4357/aa85df}, \href {https://ui.adsabs.harvard.edu/abs/2017ApJ...846..132H} {846, 132}

\bibitem[\protect\citeauthoryear{{Hillwig} \& {Gies}}{{Hillwig} \& {Gies}}{2008}]{hillwig08}
{Hillwig} T.~C.,  {Gies} D.~R.,  2008, \mn@doi [\apjl] {10.1086/587140}, \href {https://ui.adsabs.harvard.edu/abs/2008ApJ...676L..37H} {676, L37}

\bibitem[\protect\citeauthoryear{{Hynes}, {Steeghs}, {Casares}, {Charles}  \& {O'Brien}}{{Hynes} et~al.}{2003}]{hynes03}
{Hynes} R.~I.,  {Steeghs} D.,  {Casares} J.,  {Charles} P.~A.,   {O'Brien} K.,  2003, \mn@doi [\apjl] {10.1086/368108}, \href {https://ui.adsabs.harvard.edu/abs/2003ApJ...583L..95H} {583, L95}

\bibitem[\protect\citeauthoryear{{Kayama} et~al.,}{{Kayama} et~al.}{2022}]{kayama22}
{Kayama} K.,  et~al., 2022, \mn@doi [\pasj] {10.1093/pasj/psac060}, \href {https://ui.adsabs.harvard.edu/abs/2022PASJ...74.1143K} {74, 1143}

\bibitem[\protect\citeauthoryear{{Khargharia}, {Froning}  \& {Robinson}}{{Khargharia} et~al.}{2010}]{khargharia10}
{Khargharia} J.,  {Froning} C.~S.,   {Robinson} E.~L.,  2010, \mn@doi [\apj] {10.1088/0004-637X/716/2/1105}, \href {https://ui.adsabs.harvard.edu/abs/2010ApJ...716.1105K} {716, 1105}

\bibitem[\protect\citeauthoryear{{Khargharia}, {Froning}, {Robinson}  \& {Gelino}}{{Khargharia} et~al.}{2013}]{khargharia13}
{Khargharia} J.,  {Froning} C.~S.,  {Robinson} E.~L.,   {Gelino} D.~M.,  2013, \mn@doi [\aj] {10.1088/0004-6256/145/1/21}, \href {https://ui.adsabs.harvard.edu/abs/2013AJ....145...21K} {145, 21}

\bibitem[\protect\citeauthoryear{{Kubota}, {Ueda}, {Fabrika}, {Medvedev}, {Barsukova}, {Sholukhova}  \& {Goranskij}}{{Kubota} et~al.}{2010}]{kubota10}
{Kubota} K.,  {Ueda} Y.,  {Fabrika} S.,  {Medvedev} A.,  {Barsukova} E.~A.,  {Sholukhova} O.,   {Goranskij} V.~P.,  2010, \mn@doi [\apj] {10.1088/0004-637X/709/2/1374}, \href {https://ui.adsabs.harvard.edu/abs/2010ApJ...709.1374K} {709, 1374}

\bibitem[\protect\citeauthoryear{{Kuulkers} et~al.,}{{Kuulkers} et~al.}{2013}]{kuulkers13}
{Kuulkers} E.,  et~al., 2013, \mn@doi [\aap] {10.1051/0004-6361/201219447}, \href {https://ui.adsabs.harvard.edu/abs/2013A&A...552A..32K} {552, A32}

\bibitem[\protect\citeauthoryear{{Lane}, {McClintock}  \& {Remillard}}{{Lane} et~al.}{1995}]{lane95}
{Lane} B.,  {McClintock} J.~E.,   {Remillard} R.~A.,  1995, in American Astronomical Society Meeting Abstracts. p. 104.02

\bibitem[\protect\citeauthoryear{{Laycock}, {Cappallo}  \& {Moro}}{{Laycock} et~al.}{2015}]{laycock15}
{Laycock} S. G.~T.,  {Cappallo} R.~C.,   {Moro} M.~J.,  2015, \mn@doi [\mnras] {10.1093/mnras/stu2151}, \href {https://ui.adsabs.harvard.edu/abs/2015MNRAS.446.1399L} {446, 1399}

\bibitem[\protect\citeauthoryear{{Lei} et~al.,}{{Lei} et~al.}{2023}]{lei23}
{Lei} L.,  et~al., 2023, \mn@doi [arXiv e-prints] {10.48550/arXiv.2305.03408}, \href {https://ui.adsabs.harvard.edu/abs/2023arXiv230503408L} {p. arXiv:2305.03408}

\bibitem[\protect\citeauthoryear{{Liu}, {Liu}, {Bambi}  \& {Ji}}{{Liu} et~al.}{2022}]{liu22}
{Liu} Q.,  {Liu} H.,  {Bambi} C.,   {Ji} L.,  2022, \mn@doi [\mnras] {10.1093/mnras/stac616}, \href {https://ui.adsabs.harvard.edu/abs/2022MNRAS.512.2082L} {512, 2082}

\bibitem[\protect\citeauthoryear{{MacDonald} et~al.,}{{MacDonald} et~al.}{2014}]{macdonald14}
{MacDonald} R. K.~D.,  et~al., 2014, \mn@doi [\apj] {10.1088/0004-637X/784/1/2}, \href {https://ui.adsabs.harvard.edu/abs/2014ApJ...784....2M} {784, 2}

\bibitem[\protect\citeauthoryear{{Macias}, {Orosz}, {Bailyn}, {Buxton}, {Schechter}, {Remillard}, {McClintock}  \& {Steiner}}{{Macias} et~al.}{2011}]{macias11}
{Macias} P.,  {Orosz} J.~A.,  {Bailyn} C.~D.,  {Buxton} M.~M.,  {Schechter} P.~L.,  {Remillard} R.~A.,  {McClintock} J.~E.,   {Steiner} J.~F.,  2011, in American Astronomical Society Meeting Abstracts \#217. p. 143.04

\bibitem[\protect\citeauthoryear{{Maxted}, {Ruiter}, {Belczynski}, {Seitenzahl}  \& {Crocker}}{{Maxted} et~al.}{2020}]{maxted20}
{Maxted} N.~I.,  {Ruiter} A.~J.,  {Belczynski} K.,  {Seitenzahl} I.~R.,   {Crocker} R.~M.,  2020, \mn@doi [arXiv e-prints] {10.48550/arXiv.2010.15341}, \href {https://ui.adsabs.harvard.edu/abs/2020arXiv201015341M} {p. arXiv:2010.15341}

\bibitem[\protect\citeauthoryear{{Miko{\l}ajewska}, {Zdziarski}, {Zi{\'o}{\l}kowski}, {Torres}  \& {Casares}}{{Miko{\l}ajewska} et~al.}{2022}]{mikolajewska22}
{Miko{\l}ajewska} J.,  {Zdziarski} A.~A.,  {Zi{\'o}{\l}kowski} J.,  {Torres} M. A.~P.,   {Casares} J.,  2022, \mn@doi [\apj] {10.3847/1538-4357/ac6099}, \href {https://ui.adsabs.harvard.edu/abs/2022ApJ...930....9M} {930, 9}

\bibitem[\protect\citeauthoryear{{Miller-Jones} et~al.,}{{Miller-Jones} et~al.}{2021}]{miller21}
{Miller-Jones} J. C.~A.,  et~al., 2021, \mn@doi [Science] {10.1126/science.abb3363}, \href {https://ui.adsabs.harvard.edu/abs/2021Sci...371.1046M} {371, 1046}

\bibitem[\protect\citeauthoryear{{Molla}, {Chakrabarti}, {Debnath}, {Mondal}, {Jana}  \& {Chatterjee}}{{Molla} et~al.}{2015}]{molla15}
{Molla} A.~A.,  {Chakrabarti} S.~K.,  {Debnath} D.,  {Mondal} S.,  {Jana} A.,   {Chatterjee} D.,  2015, in Astronomical Society of India Conference Series. pp 119--120 (\mn@eprint {arXiv} {1603.01809}), \mn@doi{10.48550/arXiv.1603.01809}

\bibitem[\protect\citeauthoryear{{Molla}, {Debnath}, {Chakrabarti}, {Mondal}  \& {Jana}}{{Molla} et~al.}{2016}]{molla16b}
{Molla} A.~A.,  {Debnath} D.,  {Chakrabarti} S.~K.,  {Mondal} S.,   {Jana} A.,  2016, \mn@doi [\mnras] {10.1093/mnras/stw860}, \href {https://ui.adsabs.harvard.edu/abs/2016MNRAS.460.3163M} {460, 3163}

\bibitem[\protect\citeauthoryear{{Motta}, {Belloni}, {Stella}, {Mu{\~n}oz-Darias}  \& {Fender}}{{Motta} et~al.}{2014}]{motta14}
{Motta} S.~E.,  {Belloni} T.~M.,  {Stella} L.,  {Mu{\~n}oz-Darias} T.,   {Fender} R.,  2014, \mn@doi [\mnras] {10.1093/mnras/stt2068}, \href {https://ui.adsabs.harvard.edu/abs/2014MNRAS.437.2554M} {437, 2554}

\bibitem[\protect\citeauthoryear{{Orosz}}{{Orosz}}{2002}]{orosz02b}
{Orosz} J.,  2002, in X-ray Binaries in the Chandra and XMM-Newton Era (with an emphasis on Targets of Opportunity). p.~30

\bibitem[\protect\citeauthoryear{{Orosz}, {Jain}, {Bailyn}, {McClintock}  \& {Remillard}}{{Orosz} et~al.}{1998}]{orosz98}
{Orosz} J.~A.,  {Jain} R.~K.,  {Bailyn} C.~D.,  {McClintock} J.~E.,   {Remillard} R.~A.,  1998, in {Holt} S.~S.,  {Kallman} T.~R.,  eds,  American Institute of Physics Conference Series Vol. 431, Accretion processes in Astrophysical Systems: Some like it hot! - eigth AstroPhysics Conference. pp 339--342, \mn@doi{10.1063/1.55872}

\bibitem[\protect\citeauthoryear{{Orosz} et~al.,}{{Orosz} et~al.}{2001}]{orosz01}
{Orosz} J.~A.,  et~al., 2001, \mn@doi [\apj] {10.1086/321442}, \href {https://ui.adsabs.harvard.edu/abs/2001ApJ...555..489O} {555, 489}

\bibitem[\protect\citeauthoryear{{Orosz} et~al.,}{{Orosz} et~al.}{2002}]{orosz02a}
{Orosz} J.~A.,  et~al., 2002, \mn@doi [\apj] {10.1086/338984}, \href {https://ui.adsabs.harvard.edu/abs/2002ApJ...568..845O} {568, 845}

\bibitem[\protect\citeauthoryear{{Orosz} et~al.,}{{Orosz} et~al.}{2007}]{orosz07}
{Orosz} J.~A.,  et~al., 2007, \mn@doi [\nat] {10.1038/nature06218}, \href {https://ui.adsabs.harvard.edu/abs/2007Natur.449..872O} {449, 872}

\bibitem[\protect\citeauthoryear{{Orosz} et~al.,}{{Orosz} et~al.}{2009}]{orosz09}
{Orosz} J.~A.,  et~al., 2009, \mn@doi [\apj] {10.1088/0004-637X/697/1/573}, \href {https://ui.adsabs.harvard.edu/abs/2009ApJ...697..573O} {697, 573}

\bibitem[\protect\citeauthoryear{{Orosz}, {Steiner}, {McClintock}, {Torres}, {Remillard}, {Bailyn}  \& {Miller}}{{Orosz} et~al.}{2011a}]{orosz11b}
{Orosz} J.~A.,  {Steiner} J.~F.,  {McClintock} J.~E.,  {Torres} M. A.~P.,  {Remillard} R.~A.,  {Bailyn} C.~D.,   {Miller} J.~M.,  2011a, \mn@doi [\apj] {10.1088/0004-637X/730/2/75}, \href {https://ui.adsabs.harvard.edu/abs/2011ApJ...730...75O} {730, 75}

\bibitem[\protect\citeauthoryear{{Orosz}, {McClintock}, {Aufdenberg}, {Remillard}, {Reid}, {Narayan}  \& {Gou}}{{Orosz} et~al.}{2011b}]{orosz11a}
{Orosz} J.~A.,  {McClintock} J.~E.,  {Aufdenberg} J.~P.,  {Remillard} R.~A.,  {Reid} M.~J.,  {Narayan} R.,   {Gou} L.,  2011b, \mn@doi [\apj] {10.1088/0004-637X/742/2/84}, \href {https://ui.adsabs.harvard.edu/abs/2011ApJ...742...84O} {742, 84}

\bibitem[\protect\citeauthoryear{{Orosz}, {Steiner}, {McClintock}, {Buxton}, {Bailyn}, {Steeghs}, {Guberman}  \& {Torres}}{{Orosz} et~al.}{2014}]{orosz14}
{Orosz} J.~A.,  {Steiner} J.~F.,  {McClintock} J.~E.,  {Buxton} M.~M.,  {Bailyn} C.~D.,  {Steeghs} D.,  {Guberman} A.,   {Torres} M. A.~P.,  2014, \mn@doi [\apj] {10.1088/0004-637X/794/2/154}, \href {https://ui.adsabs.harvard.edu/abs/2014ApJ...794..154O} {794, 154}

\bibitem[\protect\citeauthoryear{{{\"O}zel}, {Psaltis}, {Narayan}  \& {McClintock}}{{{\"O}zel} et~al.}{2010}]{ozel10}
{{\"O}zel} F.,  {Psaltis} D.,  {Narayan} R.,   {McClintock} J.~E.,  2010, \mn@doi [\apj] {10.1088/0004-637X/725/2/1918}, \href {https://ui.adsabs.harvard.edu/abs/2010ApJ...725.1918O} {725, 1918}

\bibitem[\protect\citeauthoryear{{Parker} et~al.,}{{Parker} et~al.}{2016}]{parker16}
{Parker} M.~L.,  et~al., 2016, \mn@doi [\apjl] {10.3847/2041-8205/821/1/L6}, \href {https://ui.adsabs.harvard.edu/abs/2016ApJ...821L...6P} {821, L6}

\bibitem[\protect\citeauthoryear{{Pietsch}, {Haberl}, {Sasaki}, {Gaetz}, {Plucinsky}, {Ghavamian}, {Long}  \& {Pannuti}}{{Pietsch} et~al.}{2006}]{pietsch06}
{Pietsch} W.,  {Haberl} F.,  {Sasaki} M.,  {Gaetz} T.~J.,  {Plucinsky} P.~P.,  {Ghavamian} P.,  {Long} K.~S.,   {Pannuti} T.~G.,  2006, \mn@doi [\apj] {10.1086/504704}, \href {https://ui.adsabs.harvard.edu/abs/2006ApJ...646..420P} {646, 420}

\bibitem[\protect\citeauthoryear{{Planck Collaboration} et~al.,}{{Planck Collaboration} et~al.}{2020}]{planck18}
{Planck Collaboration} et~al., 2020, \mn@doi [\aap] {10.1051/0004-6361/201833910}, \href {https://ui.adsabs.harvard.edu/abs/2020A&A...641A...6P} {641, A6}

\bibitem[\protect\citeauthoryear{{Queiroz} et~al.,}{{Queiroz} et~al.}{2018}]{queiroz18}
{Queiroz} A.~B.~A.,  et~al., 2018, \mn@doi [\mnras] {10.1093/mnras/sty330}, \href {https://ui.adsabs.harvard.edu/abs/2018MNRAS.476.2556Q} {476, 2556}

\bibitem[\protect\citeauthoryear{{Queiroz} et~al.,}{{Queiroz} et~al.}{2023}]{queiroz23}
{Queiroz} A.~B.~A.,  et~al., 2023, \mn@doi [\aap] {10.1051/0004-6361/202245399}, \href {https://ui.adsabs.harvard.edu/abs/2023A&A...673A.155Q} {673, A155}

\bibitem[\protect\citeauthoryear{{Rodriguez}}{{Rodriguez}}{2023}]{rodriguez23}
{Rodriguez} C.~L.,  2023, \mn@doi [\apjl] {10.3847/2041-8213/acc9b6}, \href {https://ui.adsabs.harvard.edu/abs/2023ApJ...947L..12R} {947, L12}

\bibitem[\protect\citeauthoryear{{Salcido} et~al.,}{{Salcido} et~al.}{2018}]{salcido18}
{Salcido} J.,  et~al., 2018, \mn@doi [\mnras] {10.1093/mnras/sty879}, \href {https://ui.adsabs.harvard.edu/abs/2018MNRAS.477.3744S} {477, 3744}

\bibitem[\protect\citeauthoryear{{Shahbaz}, {van der Hooft}, {Charles}, {Casares}  \& {van Paradijs}}{{Shahbaz} et~al.}{1996}]{shahbaz97}
{Shahbaz} T.,  {van der Hooft} F.,  {Charles} P.~A.,  {Casares} J.,   {van Paradijs} J.,  1996, \mn@doi [\mnras] {10.1093/mnras/282.4.L47}, \href {https://ui.adsabs.harvard.edu/abs/1996MNRAS.282L..47S} {282, L47}

\bibitem[\protect\citeauthoryear{{Shang}, {Debnath}, {Chatterjee}, {Jana}, {Chakrabarti}, {Chang}, {Yap}  \& {Chiu}}{{Shang} et~al.}{2019}]{shang19}
{Shang} J.~R.,  {Debnath} D.,  {Chatterjee} D.,  {Jana} A.,  {Chakrabarti} S.~K.,  {Chang} H.~K.,  {Yap} Y.~X.,   {Chiu} C.~L.,  2019, \mn@doi [\apj] {10.3847/1538-4357/ab0c1e}, \href {https://ui.adsabs.harvard.edu/abs/2019ApJ...875....4S} {875, 4}

\bibitem[\protect\citeauthoryear{{Shaposhnikov} \& {Titarchuk}}{{Shaposhnikov} \& {Titarchuk}}{2007}]{shaposhnikov07}
{Shaposhnikov} N.,  {Titarchuk} L.,  2007, \mn@doi [\apj] {10.1086/518110}, \href {https://ui.adsabs.harvard.edu/abs/2007ApJ...663..445S} {663, 445}

\bibitem[\protect\citeauthoryear{{Shaposhnikov} \& {Titarchuk}}{{Shaposhnikov} \& {Titarchuk}}{2009}]{shaposhnikov09}
{Shaposhnikov} N.,  {Titarchuk} L.,  2009, \mn@doi [\apj] {10.1088/0004-637X/699/1/453}, \href {https://ui.adsabs.harvard.edu/abs/2009ApJ...699..453S} {699, 453}

\bibitem[\protect\citeauthoryear{{Shenar} et~al.,}{{Shenar} et~al.}{2022}]{shenar22}
{Shenar} T.,  et~al., 2022, \mn@doi [Nature Astronomy] {10.1038/s41550-022-01730-y}, \href {https://ui.adsabs.harvard.edu/abs/2022NatAs...6.1085S} {6, 1085}

\bibitem[\protect\citeauthoryear{{Sreehari}, {Iyer}, {Radhika}, {Nandi}  \& {Mandal}}{{Sreehari} et~al.}{2019a}]{sreehari19b}
{Sreehari} H.,  {Iyer} N.,  {Radhika} D.,  {Nandi} A.,   {Mandal} S.,  2019a, \mn@doi [Advances in Space Research] {10.1016/j.asr.2018.10.042}, \href {https://ui.adsabs.harvard.edu/abs/2019AdSpR..63.1374S} {63, 1374}

\bibitem[\protect\citeauthoryear{{Sreehari}, {Ravishankar}, {Iyer}, {Agrawal}, {Katoch}, {Mandal}  \& {Nandi}}{{Sreehari} et~al.}{2019b}]{sreehari19a}
{Sreehari} H.,  {Ravishankar} B.~T.,  {Iyer} N.,  {Agrawal} V.~K.,  {Katoch} T.~B.,  {Mandal} S.,   {Nandi} A.,  2019b, \mn@doi [\mnras] {10.1093/mnras/stz1327}, \href {https://ui.adsabs.harvard.edu/abs/2019MNRAS.487..928S} {487, 928}

\bibitem[\protect\citeauthoryear{{Sridhar}, {Bhattacharyya}, {Chandra}  \& {Antia}}{{Sridhar} et~al.}{2019}]{sridhar19}
{Sridhar} N.,  {Bhattacharyya} S.,  {Chandra} S.,   {Antia} H.~M.,  2019, \mn@doi [\mnras] {10.1093/mnras/stz1476}, \href {https://ui.adsabs.harvard.edu/abs/2019MNRAS.487.4221S} {487, 4221}

\bibitem[\protect\citeauthoryear{{Thompson} et~al.,}{{Thompson} et~al.}{2018}]{thompson18}
{Thompson} B.~B.,  et~al., 2018, \mn@doi [\mnras] {10.1093/mnras/stx2316}, \href {https://ui.adsabs.harvard.edu/abs/2018MNRAS.473..185T} {473, 185}

\bibitem[\protect\citeauthoryear{{Torres}, {Casares}, {Jim{\'e}nez-Ibarra}, {Mu{\~n}oz-Darias}, {Armas Padilla}, {Jonker}  \& {Heida}}{{Torres} et~al.}{2019}]{torres19}
{Torres} M.~A.~P.,  {Casares} J.,  {Jim{\'e}nez-Ibarra} F.,  {Mu{\~n}oz-Darias} T.,  {Armas Padilla} M.,  {Jonker} P.~G.,   {Heida} M.,  2019, \mn@doi [\apjl] {10.3847/2041-8213/ab39df}, \href {https://ui.adsabs.harvard.edu/abs/2019ApJ...882L..21T} {882, L21}

\bibitem[\protect\citeauthoryear{{Torres}, {Casares}, {Jim{\'e}nez-Ibarra}, {{\'A}lvarez-Hern{\'a}ndez}, {Mu{\~n}oz-Darias}, {Armas Padilla}, {Jonker}  \& {Heida}}{{Torres} et~al.}{2020}]{torres20}
{Torres} M.~A.~P.,  {Casares} J.,  {Jim{\'e}nez-Ibarra} F.,  {{\'A}lvarez-Hern{\'a}ndez} A.,  {Mu{\~n}oz-Darias} T.,  {Armas Padilla} M.,  {Jonker} P.~G.,   {Heida} M.,  2020, \mn@doi [\apjl] {10.3847/2041-8213/ab863a}, \href {https://ui.adsabs.harvard.edu/abs/2020ApJ...893L..37T} {893, L37}

\bibitem[\protect\citeauthoryear{{Val-Baker}, {Norton}  \& {Negueruela}}{{Val-Baker} et~al.}{2007}]{val07}
{Val-Baker} A.~K.~F.,  {Norton} A.~J.,   {Negueruela} I.,  2007, in {di Salvo} T.,  {Israel} G.~L.,  {Piersant} L.,  {Burderi} L.,  {Matt} G.,  {Tornambe} A.,   {Menna} M.~T.,  eds,  American Institute of Physics Conference Series Vol. 924, The Multicolored Landscape of Compact Objects and Their Explosive Origins. pp 530--533 (\mn@eprint {arXiv} {1608.01187}), \mn@doi{10.1063/1.2774906}

\bibitem[\protect\citeauthoryear{{Zampieri}, {Shapiro}  \& {Colpi}}{{Zampieri} et~al.}{1998}]{zampieri98}
{Zampieri} L.,  {Shapiro} S.~L.,   {Colpi} M.,  1998, \mn@doi [\apjl] {10.1086/311511}, \href {https://ui.adsabs.harvard.edu/abs/1998ApJ...502L.149Z} {502, L149}

\bibitem[\protect\citeauthoryear{{Zwitter}, {Calvani}  \& {D'Odorico}}{{Zwitter} et~al.}{1991}]{zwitter91}
{Zwitter} T.,  {Calvani} M.,   {D'Odorico} S.,  1991, \aap, \href {https://ui.adsabs.harvard.edu/abs/1991A&A...251...92Z} {251, 92}

\makeatother
\end{thebibliography}

\appendix\section{Comments on individual sources}
Here we comment on adopted values of ages and masses of individual BHs. We skip objects for which we found only one source for their mass or age in the literature. Final values and adopted sources for all objects are listed in Tables~\ref{mlade} and ~\ref{stare}. 

As mentioned above, BHs were divided into two groups. Here we comment why some BHs are young. Individual ages of older BHs are discussed in Sec.~2.2. 

\subsection{Young BHs}
\label{A1}
\subsubsection{Ages}
Young BHs are the ones with ages smaller than 0.5 Gyr. Below we justify why certain BHs belong to this group.

The first clue that the system should be young is the age of its companion. So all systems with type O or B companion were considered young. 

GRO J1655--40 was sorted among young systems because the age is known to be 40.55 Myr, since it is very likely to be a part of a young open cluster NGC 6242 \citep{combi07}. Although there is some scepticism about this connection, we have another hint that the X-ray binary is young, namely, its chemical abundance. Lithium abundance and effective temperature are consistent with an age of 1 to 3 Gyr \citep{gonzalez08}, suggesting that the system is  young. 

The system MAXI J1535--571 is connected to the supernova remnant G323.7--1.0 \citep{maxted20}. Since supernova remnants that can be observed are considered to be relatively young, we can count this system as a young one. There is another such binary, namely SS 433, which is connected to a remnant W50 \citep[see, e.g.][]{kayama22}. 

SN 1997D is a supernova, which differs from all other sources that we consider in this analysis, because it is not even a part of a binary. We can still determine the mass of this BH by knowing how much mass was ejected and modelling the supernova. Whether it is a neutron star or a BH is still under debate, so we consider this object only in the discussion section \citep{zampieri98}.

\subsubsection{Masses}

Here, we discuss masses of objects for which more than one estimate of BH mass exists in the literature. For the mass of BH in GRO J1655--40 we adopted $5.31 \pm 0.07\, \mathrm{M}_\odot$ determined with quasi-periodic oscillations (QPO) by \citet{motta14}. This new method of determining the mass of a BH assumes that these oscillations are well-described by the relativistic precession model. It agrees with the mass determined by a well-established method of light-curve fitting by \citet{beer02}. 

For NGC~300 X--1 multiple observations  agree with each other \citep{crowther10,binder19,binder21}. All of them use spectroscopic methods to determine the dynamical parameters of a Wolf-Rayet companion star. We adopted the third value determined from the newest data and having the smallest error. 

Multiple measurements of the mass of Cyg X--1 have been obtained by various methods. Here we discuss only the latest three measurements. We adopted the mass based on complete dynamical system modelling \citep{orosz11a}. It  agrees with that determined by \citet{miller21}, although it does not agree with measurements with quasi-periodic oscillation, which derive $8.7\pm 0.8 \,\mathrm{M}_\odot$  \citep{shaposhnikov07}. 

For XTE J1550--564 we adopted the mass of the BH of $9.10 \pm 0.61\,\mathrm{M}_\odot$ by \citet{orosz11b}, determined by dynamical measurements. It agrees with masses from \citet{orosz02a} and \citet{shaposhnikov09}, who use the scaling technique for BH mass determination based on a correlation between the spectral index and quasi-periodic oscillations. 

The BH mass adopted for LMC X--1 was $9.49$ to $10.79 M_\odot$ from modelling with a relativistic accretion disc model \citep{bhuvana21}. This measurement is  consistent with the results of \citet{abubekerov16}. 

For V4641~Sgr there are three masses in literature that only partially agree with each other, $6.82$ to $7.42\,\mathrm{M}_\odot$ \citep{orosz02b}, $7.1$ to $9.5 \,\mathrm{M}_\odot$ \citep{goranskij03} and $6.4 \pm 0.6 \,\mathrm{M}_\odot$ \citep{macdonald14}. Since they use similar methods (determining orbital parameters of accretion disk), we adopted the third, which uses the newest measurements and has a more advanced discussion on inclination and radial velocities. 

For LMC X--3 there are two dynamical measurements of the motion of the donor star, yielding the BH mass of $9.5$ to $13.6 \,\mathrm{M}_\odot$ \citep{val07} and $6.98 \pm 0.56 \,\mathrm{M}_\odot$ \citep{orosz14}. An estimate by modelling with a relativistic accretion disc gives $5.5$ to $6.6 \,\mathrm{M}_\odot$ \citep{bhuvana21}. We adopt the newest  dynamical measurement, since it is not model-dependent and agrees with the second quoted above. 

SS 433 sparked a discussion in the literature. Disagreement about the orbital velocity of the companion gave either a BH mass of about $4 \,\mathrm{M}_\odot$ \citep{hillwig08,kubota10} or of about $10$ to $15 \,\mathrm{M}_\odot$ \citep{blundell08}. However, the measurements yielding about $4 \,\mathrm{M}_\odot$ may originate in a circumbinary disk \citep{zwitter91}, which would imply that the higher mass estimate is realistic  \citep{bowler10}. In our analysis, we adopt the mass by \citet{bowler18}, because he uses emission spectra of the circumbinary disk together with an independent method of investigating the stability of the binary orbit, with both methods yielding consistent results. Additionally, the recent results by \citet{chere19} and \citet{chere21} agree with this mass.

There is some scepticism about the mass of BH in MAXI J1535--571. No dynamical mass measurement from optical observations for this system is available, so measurements are all based on fits to the X-ray spectrum and are therefore model dependent. In 2019,  three different precise measurements were obtained. \citet{shang19} obtained  $8.8^{+1.2}_{-1.1} \,\mathrm{M}_\odot$, \citet{sreehari19a} derived a BH mass of $5.14$ to $7.83 \,\mathrm{M}_\odot$, and \citet{sridhar19} claimed it to be $10.39^{+0.61}_{-0.62} \,\mathrm{M}_\odot$. Each of them used a different model. The advantage of the last model is that it gives the distance estimate of $5.4^{+1.8}_{-1.1}$~kpc which is close to a newer estimate of  $4.1^{+0.6}_{-0.5}$~kpc \citep{Chauhan19}. It is also the only model which takes the dimensionless spin parameter of the BH \citep[which is supposed to be high, see][]{dong22,liu22} into account. There is an  additional recent measurement of the BH mass by \citet{chand22} which gives $m_{BH}$ about $8.5$ to $16 \,\mathrm{M}_\odot$, while modelling of a supernova connected to this binary system favours the mass of  $23$ to $35 \,\mathrm{M}_\odot$ \citep{maxted20}. Considering this, we used the BH mass estimate by \citet{sridhar19}, but we extended the errorbars to $\pm 2\,\mathrm{M}_\odot$.

\subsection{Other BHs}
\label{A2}
For Gaia BH1 we adopted the mass $9.32^{+0.22}_{-0.21} M_\odot$  by \citet{chakrabarti23} because they fit all available data which include also the ones used by \citep{el-badry23a} who derived  $9.62 \pm 0.18 M_\odot$. However, for the age of the object, we preferred the value by \citet{andrae23}, because it was the only analysis focusing on this quantity.

For XTE J1118--480, the masses agree with each other, ranging between $6.9$ and $8.5 \,\mathrm{M}_\odot$ \citep{gelino06,khargharia13,chatterjee19,gonzalez12}. All were obtained by fitting an accretion disk with various models. We adopted the last because it uses the newest updated data set. 

For the mass of the BH in MAXI J1659--152, we found two estimates \citep{molla15,molla16b}. They agree with each other. The first was determined only by the Two Component Advective Flow (TCAF) model, while the second also considered the measurements of quasi-periodic oscillation, so we opted for this one. 

There are four different measurements of the mass of BH in MAXI J1820+070 \citep{torres19,atri20,torres20,chakraborty20}. All of them fit the light-curves of an accretion disk but are based on different datasets. They obtain BH masses of $7$ to $8 \,\mathrm{M}_\odot$, $9.2 \pm 1.3 \,\mathrm{M}_\odot$, $5.73$ to $8.34 \,\mathrm{M}_\odot$, and $6.7$ to $13.9 \,\mathrm{M}_\odot$, respectively. Since these measurements are all independent, we decided to use the fourth, which uses the most diverse data and obtains a confidence interval which includes all masses from other datasets. We note that \citet{mikolajewska22} obtain $6.75^{+0.64}_{0.46}$~M$_\odot$ for this object, but we feel that a poorly known metallicity could inflate the error-bars so we keep the $6.7$ to $13.9 \,\mathrm{M}_\odot$ mentioned above. Note that a lower mass would strengthen our results on rejecting $n=3$ even more. 

For GX 339--4, there are two values determined by quasi periodic-oscillations, namely $ 6.0^{+3.4}_{-1.45} \,\mathrm{M}_\odot$ \citep{bezayiff06} and $7.5 \pm 0.8 \,\mathrm{M}_\odot$ \citep{chen11}). There is also a measurement of the dynamical behaviour of the donor star which yields $2.3$ to $9.5 \,\mathrm{M}_\odot$ \citep{heida17}. Another estimate is the measurement by \citet{parker16} using spectral fitting given as $9.0^{+1.6}_{-1.2} \,\mathrm{M}_\odot$ and the estimate by \citet{sreehari19b} which claims $8.28$ to $11.89 \,\mathrm{M}_\odot$ from three different methods. The first method they use is broadband spectral modelling, the second is direct modelling of the time evolution of QPO frequencies, and the third is scaling a mass dependent parameter from an empirical model of the photon index - QPO correlation. We adopt the mass by \citet{sreehari19b} because it agrees with most of the other, and uses diverse approaches which give consistent results. 

For A0620--00, two similar measurements are used for fitting of the accretion disk  \citep{lane95,cantrell10}. They agree with each other. We adopted the second since it uses an extended sample of data.

\bsp	
\label{lastpage}
\end{document}